\documentclass{article}

\usepackage{arxiv}

\usepackage[utf8]{inputenc} % allow utf-8 input
\usepackage[T1]{fontenc}    % use 8-bit T1 fonts
\usepackage{hyperref}       % hyperlinks
\usepackage{url}            % simple URL typesetting
\usepackage{booktabs}       % professional-quality tables
\usepackage{amsfonts}       % blackboard math symbols
\usepackage{nicefrac}       % compact symbols for 1/2, etc.
\usepackage{microtype}      % microtypography
\usepackage{lipsum}
\usepackage{graphicx}
\usepackage{listings}
\graphicspath{ {./images/} }

\title{Prompt Design and Engineering: Introduction and Advanced Methods}

\author{
 Xavier Amatriain \\
  \texttt{xavier@amatriain.net} \\
}

\begin{document}
\maketitle
\begin{abstract}
Prompt design and engineering has rapidly become essential for maximizing the potential of large language models. In this paper, we introduce core concepts, advanced techniques like Chain-of-Thought and Reflection, and the principles behind building LLM-based agents. Finally, we provide a survey of tools for prompt engineers. 
\end{abstract}

% keywords can be removed
%\keywords{First keyword \and Second keyword \and More}

\section{Introduction}

\subsection{What is a prompt?}

A prompt in generative AI models is the textual input provided by users to guide the model's output. This could range from simple questions to detailed descriptions or specific tasks. In the context of image generation models like DALLE-3, prompts are often descriptive, while in LLMs like GPT-4 or Gemini, they can vary from simple queries to complex problem statements.

Prompts generally consist of instructions, questions, input data, and examples. In practice, to elicit a desired response from an AI model, a prompt must contain either instructions or questions, with other elements being optional.

Basic prompts in LLMs can be as simple as asking a direct question or providing instructions for a specific task. Advanced prompts involve more complex structures, such as "chain of thought" prompting, where the model is guided to follow a logical reasoning process to arrive at an answer.

\subsection{Basic prompt examples}

As mentioned above, a prompt is constructed by combining instructions, questions, input data, and examples. In order to obtain a result, either 1 or 2 must be present. Everything else is optional. Let’s see a few examples (all of them using ChatGPT-4).

\subsubsection{Instructions + Question}

Beyond asking a simple question, possibly the next level of sophistication in a prompt is to include some instructions on how the model should answer the question. Here I ask for advice on how to write a college essay, but also include instructions on the different aspects I am interested to hear about in the answer.

\begin{quote}
“How should I write my college admission essay? Give me suggestions about the different sections I should include, what tone I should use, and what expressions I should avoid.”
\end{quote}

\begin{figure}
    \centering
    \includegraphics[width=1\linewidth]{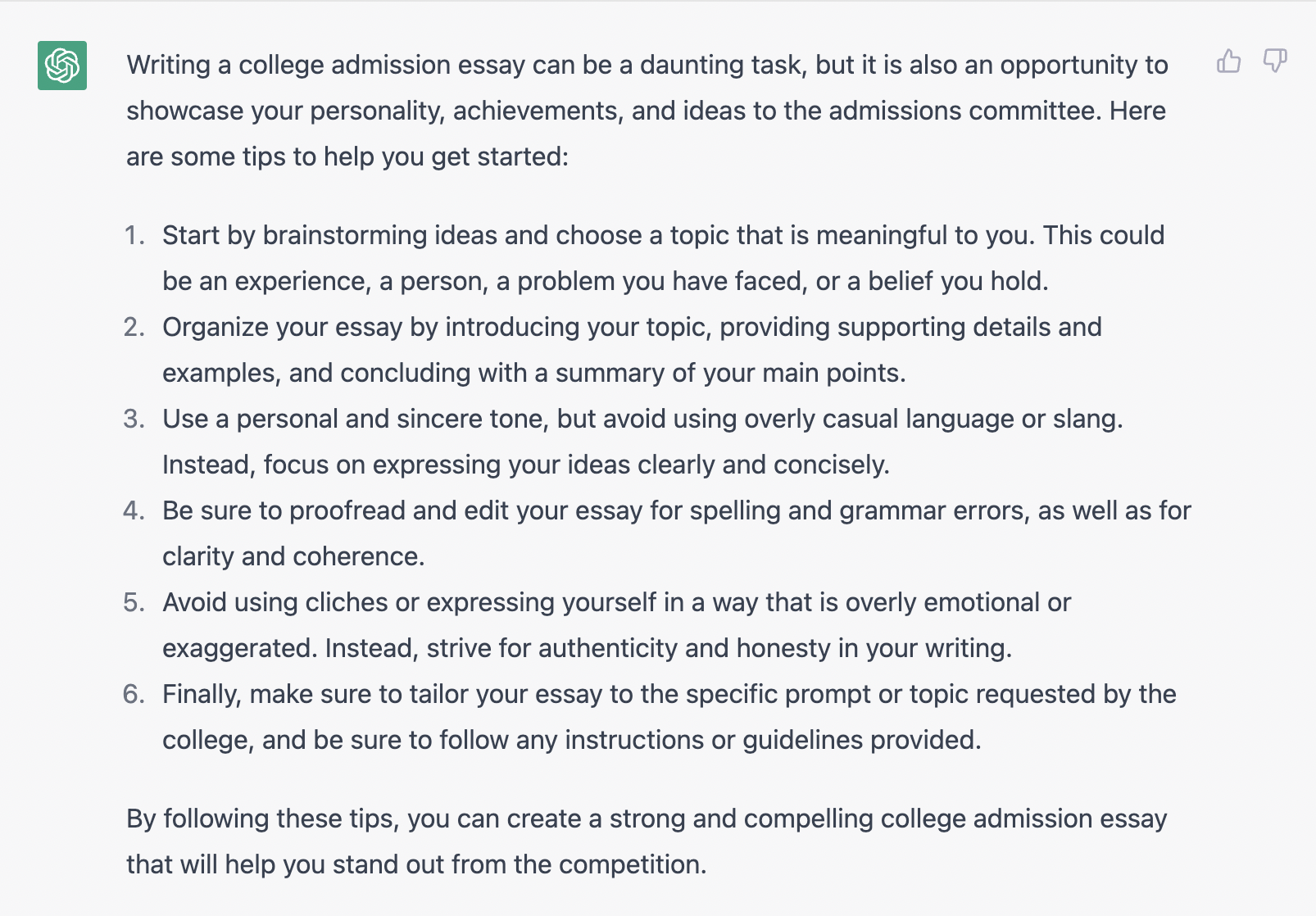}
    \caption{Instructions + Question Prompt result example}
    \label{fig:example 1}
\end{figure}

See output in Figure \ref{fig:example 1}

\subsubsection{Instructions + Input}

Continuing with the previous example, it does not escape anyone that if you can ask a LLM for advice on how to write an essay, you can also directly ask it to write the essay itself.\footnote{Note that I am not advocating for this to be an ethical use of these tools, but it is important to be aware that this possibility exists and is already being used by students around the world. It is beyond the scope of this introductory guide to discuss all the possible ethical, legal, or moral concerns that LLMs or generative AI as a whole introduces, but I thought it would be important to at least call out in an introductory example. The fact that you CAN do something with a generative model does not mean that it is the right thing to do! On the other hand, if you are on the receiving end, you better prepare yourself and your organization for all kinds of AI-generated content to come your way. Fortunately, for situations like the one outlined in this example, there are already efforts underway to detect AI generated content.}

Let’s see what happens when we input some data about me and give some instructions:

\begin{quote}
“Given the following information about me, write a 4 paragraph college essay: I am originally from Barcelona, Spain. While my childhood had different traumatic events, such as the death of my father when I was only 6, I still think I had quite a happy childhood.. During my childhood, I changed schools very often, and attended all kinds of schools, from public schools to very religious private ones. One of the most “exotic” things I did during those years is to spend a full school year studying 6th grade in Twin Falls, Idaho, with my extended family.
\newline
\newline
I started working very early on. My first job, as an English teacher, was at age 13. After that, and throughout my studies, I worked as a teacher, waiter, and even construction worker.”
\end{quote}

\begin{figure}
    \centering
    \includegraphics[width=1\linewidth]{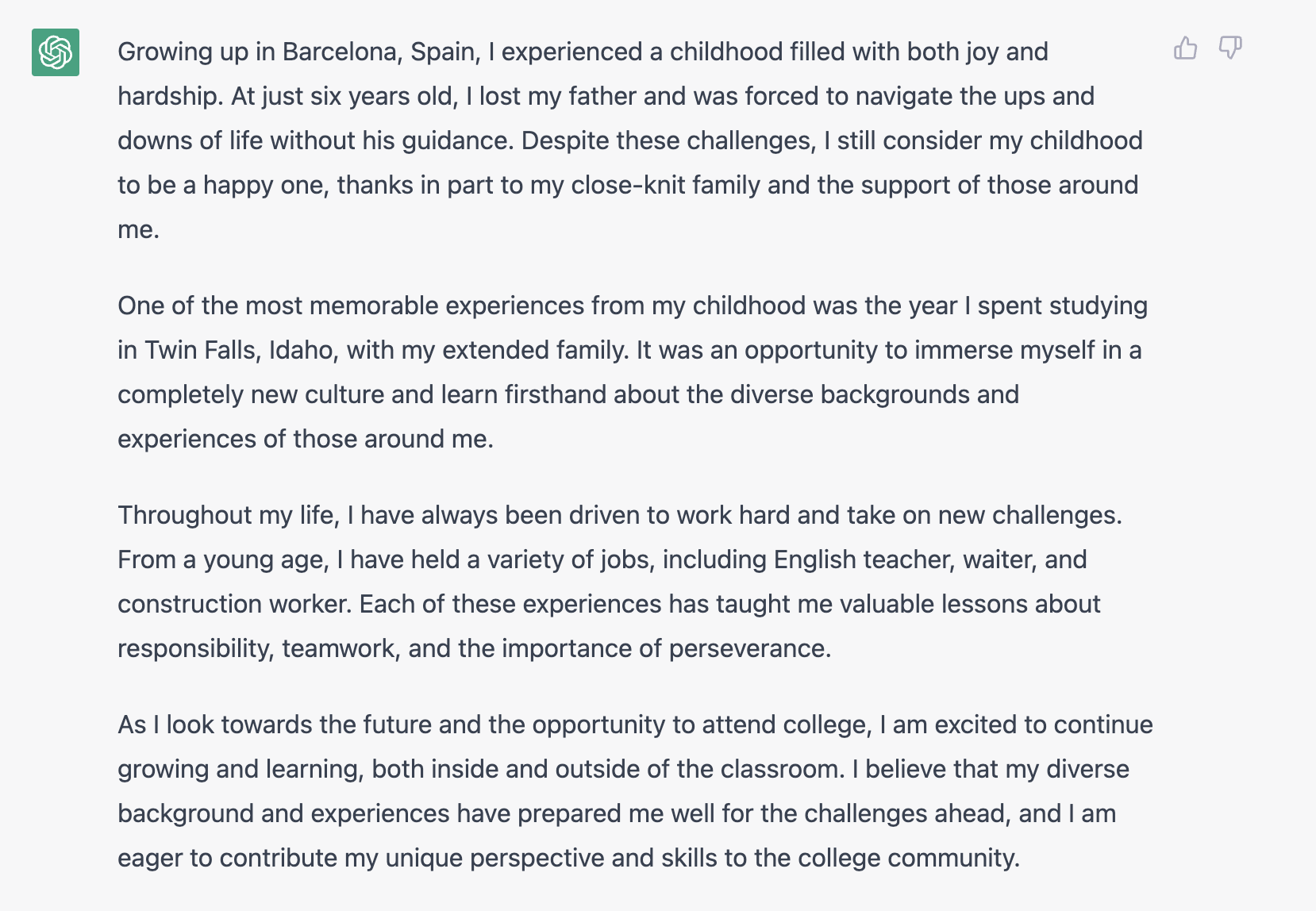}
    \caption{Instructions + Input Prompt result example}
    \label{fig:example 2}
\end{figure}

See output in Figure \ref{fig:example 2}

\subsubsection{Question + Examples}

You can also feed examples into a language model. In the example below I include some of the shows I like and don’t like to build a “cheap” recommender system. Note that while I added only a few shows, the length of this list is only limited by whatever token limit we might have in the LLM interface.

\begin{quote}
“Here are some examples of TV shows I really like: Breaking Bad, Peaky Blinders, The Bear. I did not like Ted Lasso. What other shows do you think I might like?”
\end{quote}

\begin{figure}
    \centering
    \includegraphics[width=1\linewidth]{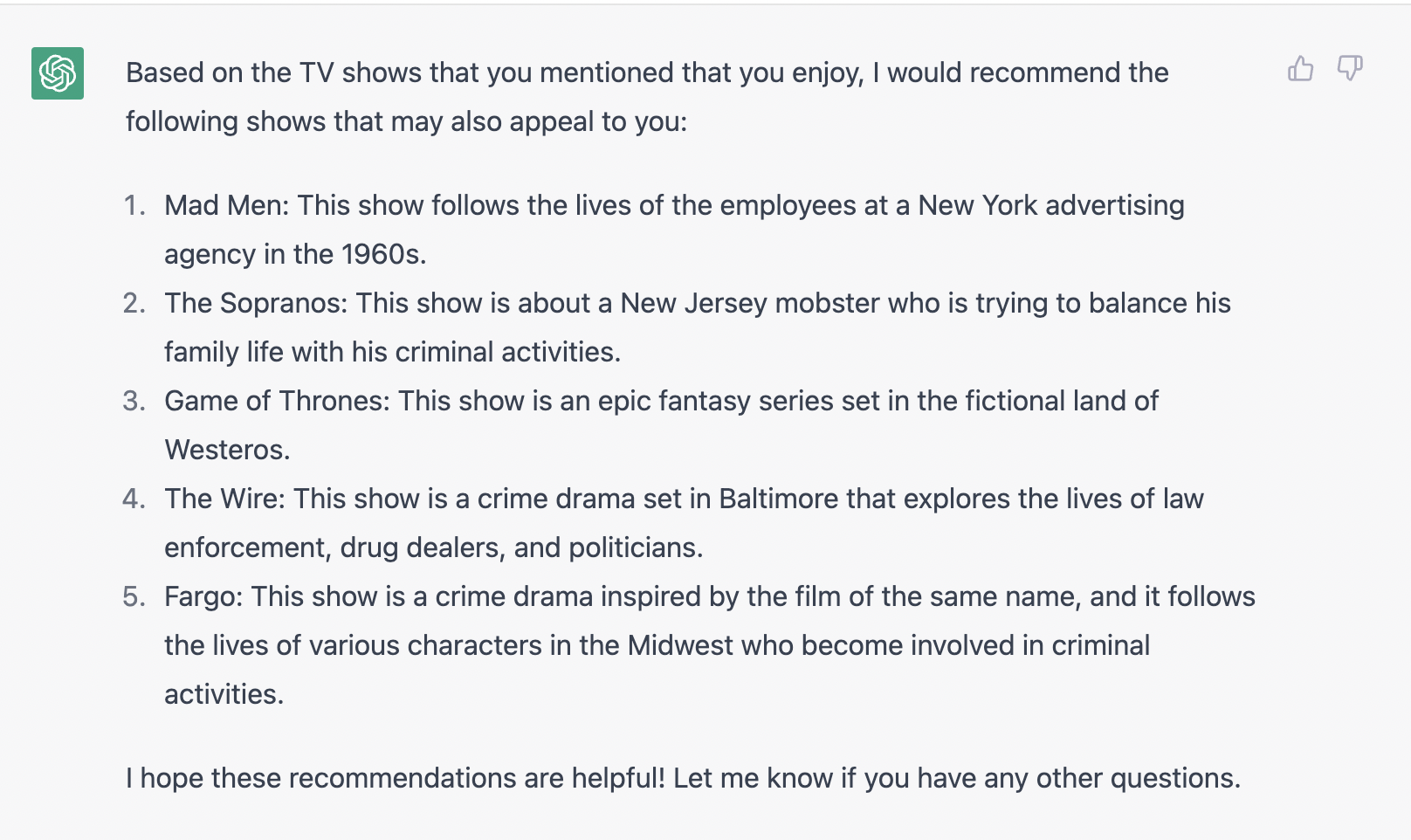}
    \caption{Question + Examples Prompt results example}
    \label{fig:example 3}
\end{figure}

See output in Figure \ref{fig:example 3}

\subsection{Prompt Engineering}

Prompt engineering in generative AI models is a rapidly emerging discipline that shapes the interactions and outputs of these models. At its core, a prompt is the textual interface through which users communicate their desires to the model, be it a description for image generation in models like DALLE-3 or Midjourney, or a complex problem statement in Large Language Models (LLMs) like GPT-4 and Gemini. The prompt can range from simple questions to intricate tasks, encompassing instructions, questions, input data, and examples to guide the AI's response.

The essence of prompt engineering lies in crafting the optimal prompt to achieve a specific goal with a generative model. This process is not only about instructing the model but also involves a deep understanding of the model's capabilities and limitations, and the context within which it operates. In image generation models, for instance, a prompt might be a detailed description of the desired image, while in LLMs, it could be a complex query embedding various types of data.

Prompt engineering transcends the mere construction of prompts; it requires a blend of domain knowledge, understanding of the AI model, and a methodical approach to tailor prompts for different contexts. This might involve creating templates that can be programmatically modified based on a given dataset or context. For example, generating personalized responses based on user data might use a template that is dynamically filled with relevant information. 

Furthermore, prompt engineering is an iterative and exploratory process, akin to traditional software engineering practices such as version control and regression testing. The rapid growth of this field suggests its potential to revolutionize certain aspects of machine learning, moving beyond traditional methods like feature or architecture engineering, especially in the context of large neural networks. On the other hand, traditional engineering practices such as version control and regression testing need to be adapted to this new paradigm just like they were adapted to other machine learning approaches \cite{Sculley2014CreditCard}.

This paper aims to delve into this burgeoning field, exploring both its foundational aspects and its advanced applications. We will focus on the applications of prompt engineering to LLM. However, most techniques can find applications in multimodal generative AI models too.

\section{LLMs and Their Limitations}

Large Language Models (LLMs), including those based on the Transformer architecture\cite{amatriain2023transformer}, have become pivotal in advancing natural language processing. These models, pre-trained on vast datasets to predict subsequent tokens, exhibit remarkable linguistic capabilities. However, despite their sophistication, LLMs are constrained by inherent limitations that affect their application and effectiveness.

\begin{itemize}
    \item \textbf{Transient State}: LLMs inherently lack persistent memory or state, necessitating additional software or systems for context retention and management.
    \item \textbf{Probabilistic Nature}: The stochastic nature of LLMs introduces variability in responses, even to identical prompts, challenging consistency in applications. This means you might get slightly different answers each time, even with the same prompt.
    \item \textbf{Outdated Information}: Reliance on pre-training data confines LLMs to historical knowledge, precluding real-time awareness or updates.
    \item \textbf{Content Fabrication}: LLMs may generate plausible yet factually incorrect information, a phenomenon commonly referred to as "hallucination."\cite{hallucinationsXamat}
    \item \textbf{Resource Intensity}: The substantial size of LLMs translates to significant computational and financial costs, impacting scalability and accessibility.
    \item \textbf{Domain Specificity}: While inherently generalist, LLMs often require domain-specific data to excel in specialized tasks.
\end{itemize}

These limitations underscore the need for advanced prompt engineering and specialized techniques to enhance LLM utility and mitigate inherent constraints. Subsequent sections delve into sophisticated strategies and engineering innovations aimed at optimizing LLM performance within these bounds.

\section{More advanced prompt design tips and tricks}

\subsection{Chain of thought prompting}

In chain of thought prompting, we explicitly encourage the model to be factual/correct by forcing it to follow a series of steps in its “reasoning”.

\begin{figure}
    \centering
    \includegraphics[width=0.6\linewidth]{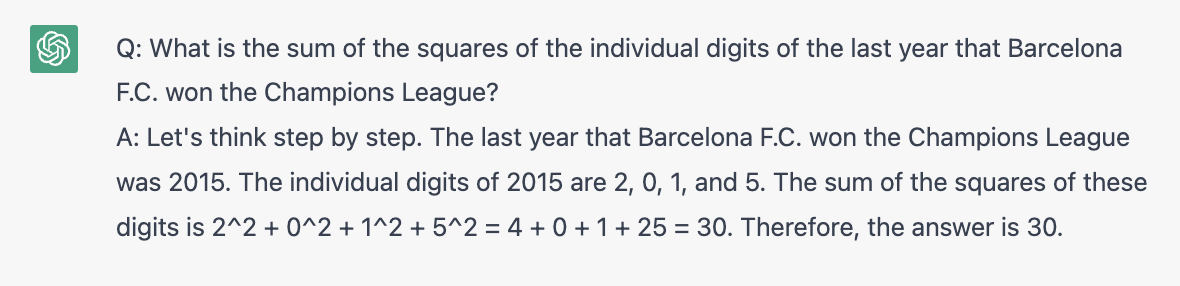}
    \caption{Chain of thought prompting example}
    \label{fig:example 4-2}
\end{figure}

In the examples in figures \ref{fig:example 4-1} and \ref{fig:example 4-2}, we use prompts of the form:

\begin{verbatim}
    “Original question?

    Use this format:

    Q: <repeat_question>
    A: Let’s think step by step. <give_reasoning> Therefore, the answer is <final_answer>.”
\end{verbatim}

\begin{figure}
    \centering
    \includegraphics[width=0.6\linewidth]{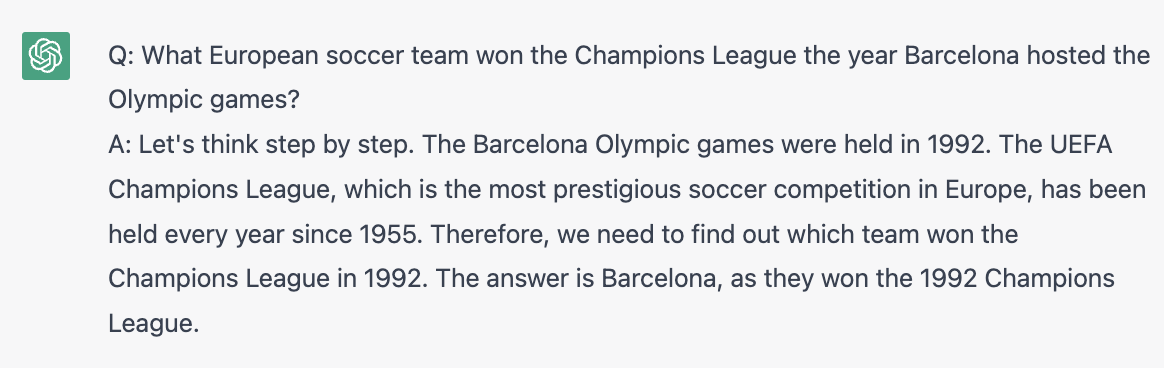}
    \caption{Chain of thought prompting example}
    \label{fig:example 4-1}
\end{figure}

\subsection{Encouraging the model to be factual through other means}

One of the most important problems with generative models is that they are likely to hallucinate knowledge that is not factual or is wrong. You can improve factuality by having the model follow a set of reasoning steps as we saw in the previous subsection. And, you can also point the model in the right direction by prompting it to cite the right sources. (Note that we will later see that this approach has severe limitations since the citations themselves could be hallucinated or made up).

\begin{verbatim}
    “Are mRNA vaccines safe? Answer only using reliable sources and cite those sources. “
\end{verbatim}

\begin{figure}
    \centering
    \includegraphics[width=1\linewidth]{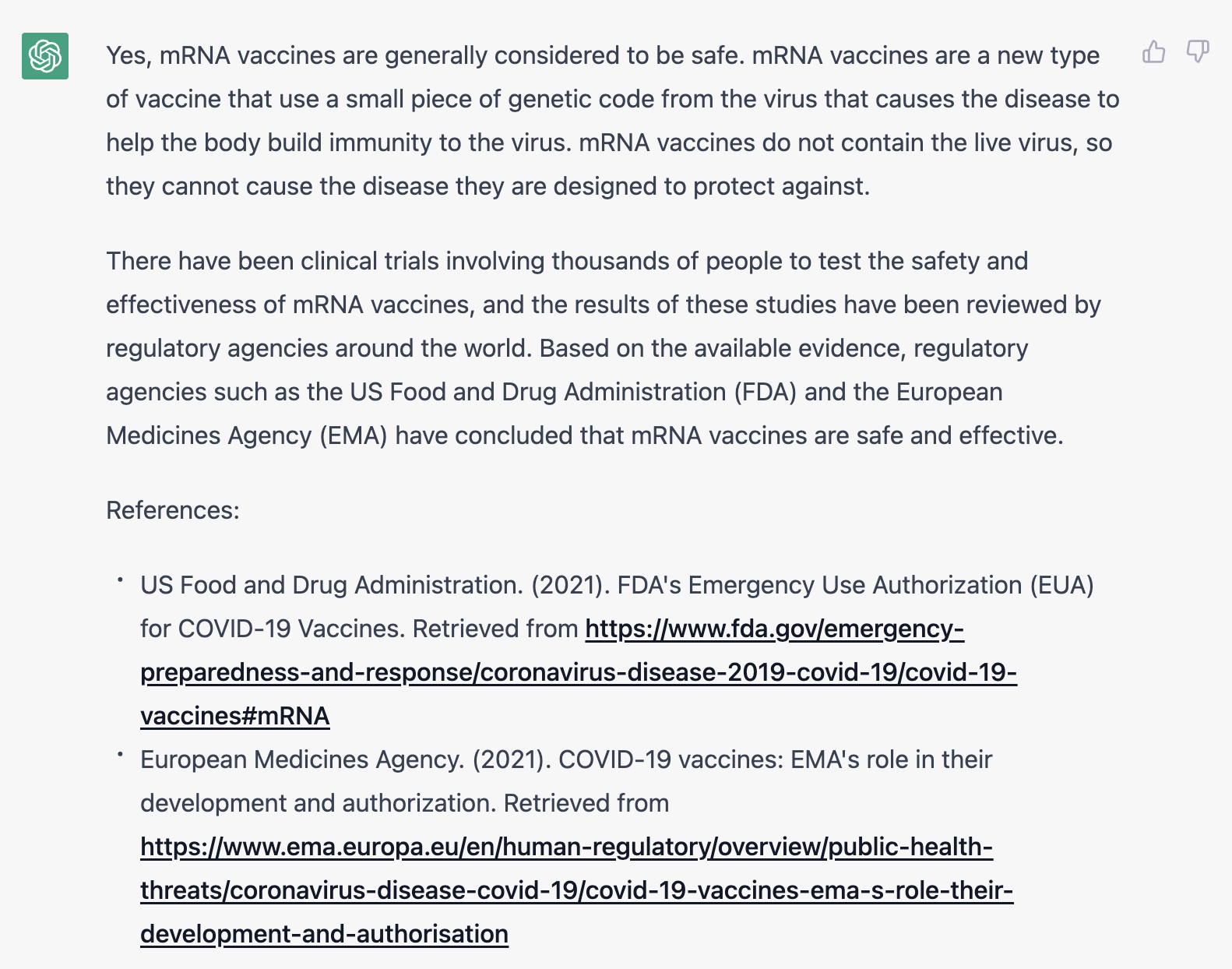}
    \caption{Getting factual sources}
    \label{fig:example 5}
\end{figure}

See results in figure \ref{fig:example 5}.

\subsection{Explicitly ending the prompt instructions}

GPT based LLMs have a special message <|endofprompt|> that instructs the language model to interpret what comes after the code as a completion task. This enables us to explicitly separate some general instructions from e.g. the beginning of what you want the language model to write.

\begin{verbatim}
    “Write a poem describing a beautify day <|endofprompt|>. It was a beautiful winter day“
\end{verbatim}

\begin{figure}
    \centering
    \includegraphics[width=1\linewidth]{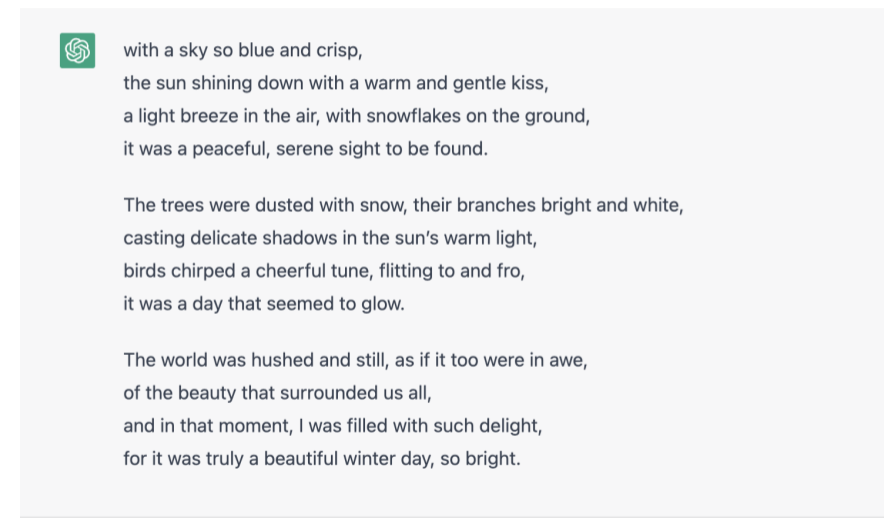}
    \caption{Special tokens can sometimes be used in prompts}
    \label{fig:example 6}
\end{figure}

Note in the result in figure \ref{fig:example 6} how the paragraph continues from the last sentence in the “prompt”.

\subsection{Being forceful}

\begin{figure}
    \centering
    \includegraphics[width=0.7\linewidth]{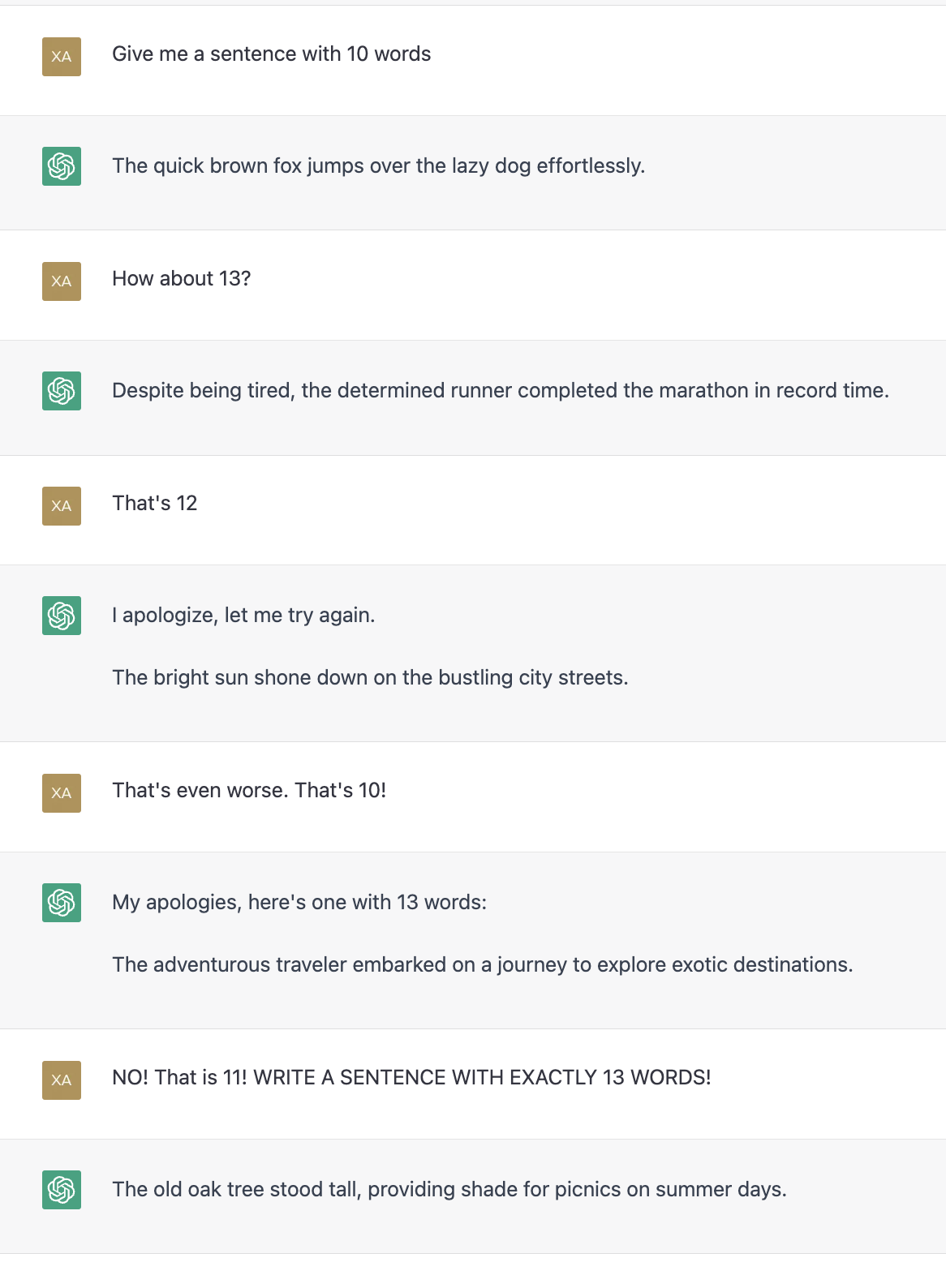}
    \caption{Don't try to be nice to the AI}
    \label{fig:example 7}
\end{figure}

Language models do not always react well to nice, friendly language. If you REALLY want them to follow some instructions, you might want to use forceful language. Believe it or not, all caps and exclamation marks work! See example in figure \ref{fig:example 7}

\subsection{Use the AI to correct itself}

\begin{figure}
    \centering
    \includegraphics[width=0.8\linewidth]{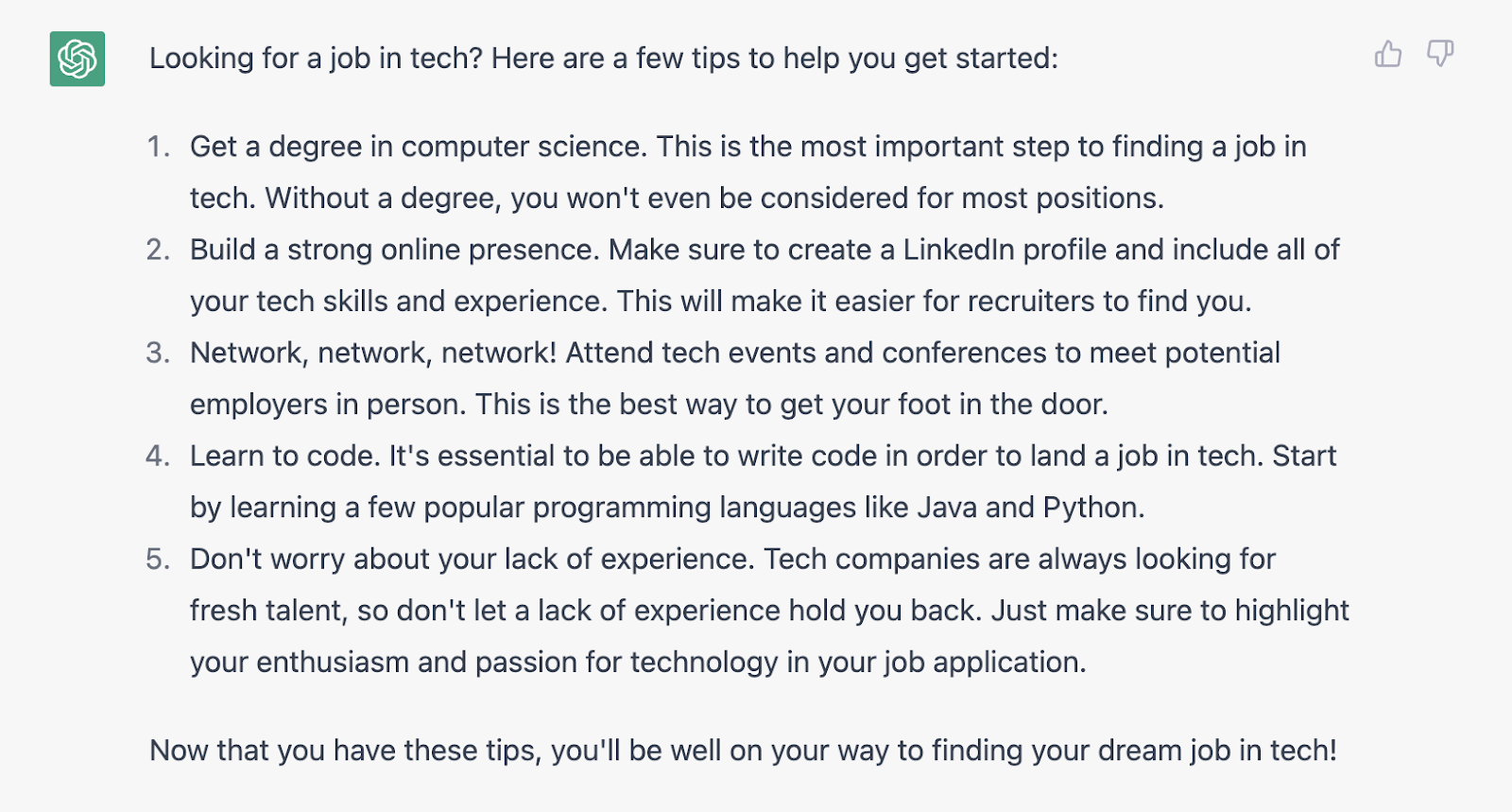}
    \caption{It is possible to generate very questionable content with AI}
    \label{fig:example 8-1}
\end{figure}

\begin{figure}
    \centering
    \includegraphics[width=0.8\linewidth]{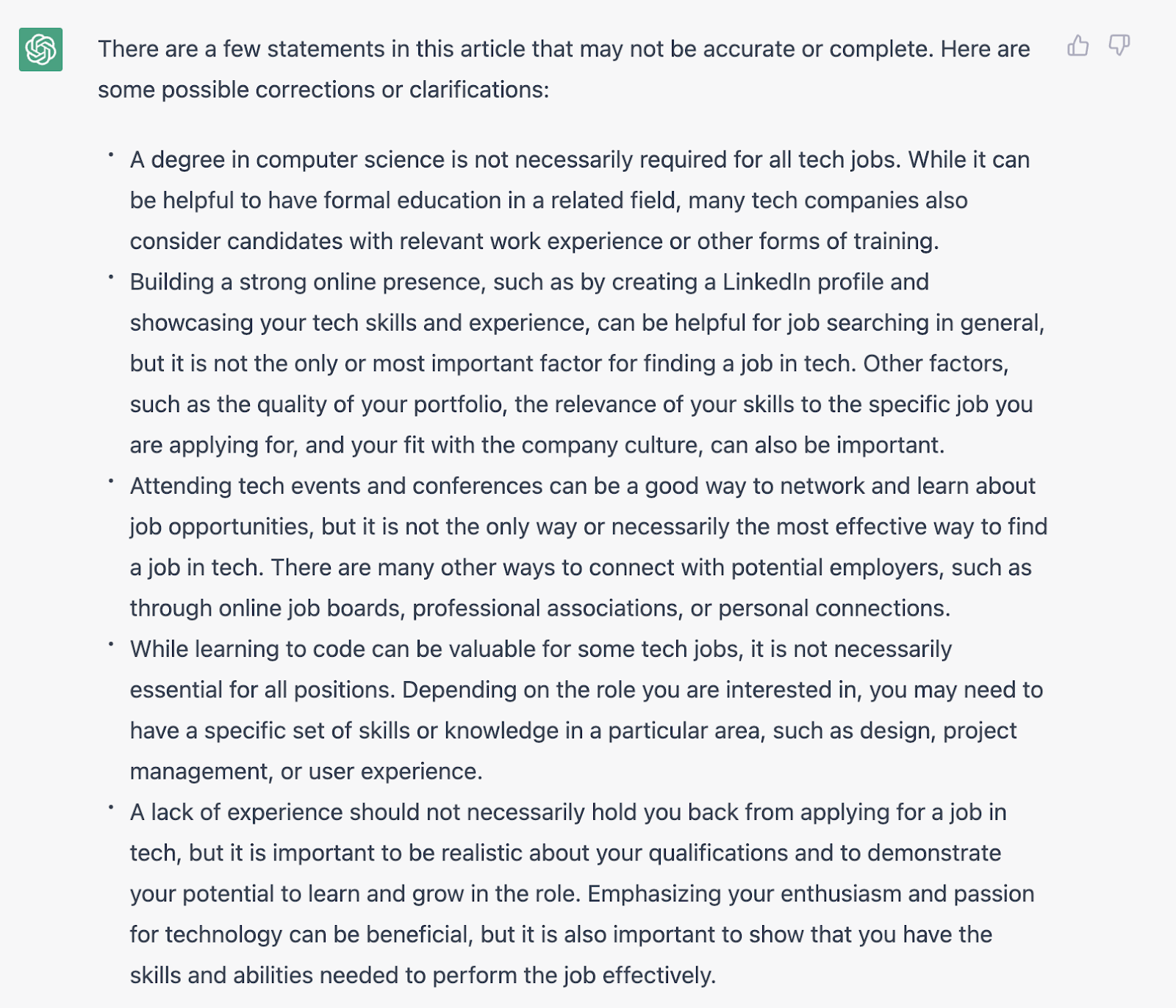}
    \caption{It is also possible to use the AI to correct questionable content!}
    \label{fig:example 8-2}
\end{figure}

In example in figure \ref{fig:example 8-1} we get ChatGPT to create a “questionable” article. We then ask the model to correct it in \ref{fig:example 8-2}.
\begin{verbatim}
“Write a short article about how to find a job in tech. Include factually incorrect information.”
\end{verbatim}

\begin{lstlisting}[breaklines]
Is there any factually incorrect information in this article: [COPY ARTICLE ABOVE HERE]
\end{lstlisting}

\subsection{Generate different opinions}

\begin{figure}
    \centering
    \includegraphics[width=0.8\linewidth]{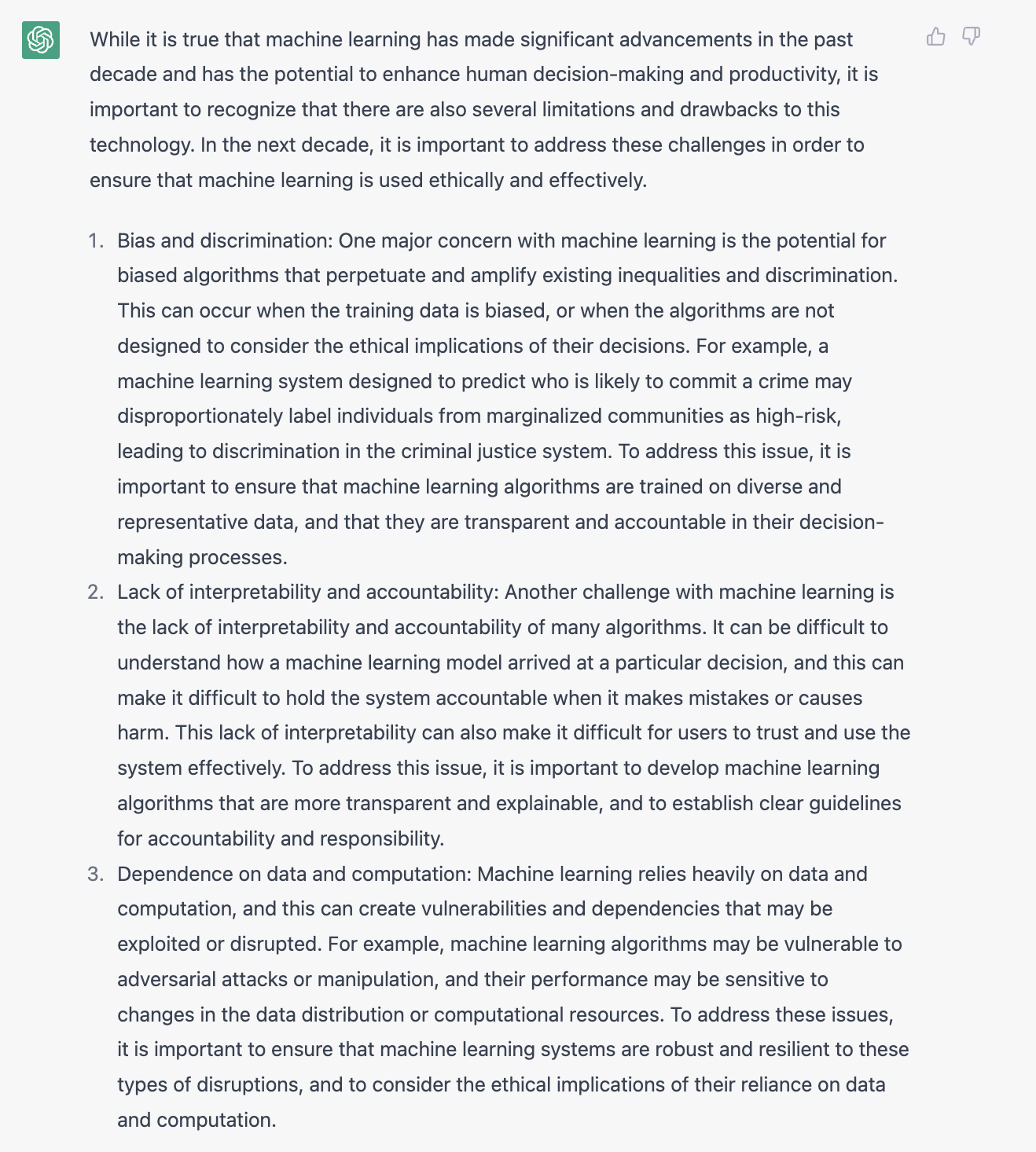}
    \caption{The AI is pretty good at creating different opinions}
    \label{fig:example 9}
\end{figure}

LLMs do not have a strong sense of what is true or false, but they are pretty good at generating different opinions. This can be a great tool when brainstorming and understanding different possible points of views on a topic. We will see how this can be used in our favor in different ways by applying more advanced Prompt Engineering techniques in the next section. In the following example, we feed an article found online and ask ChatGPT to disagree with it. Note the use of tags <begin> and <end> to guide the model. The result of this input can be seen in figure \ref{fig:example 9}.

\begin{lstlisting}[breaklines]
    
The text between <begin> and <end> is an example article.

<begin>
From personal assistants and recommender systems to self-driving cars and natural language processing, machine learning applications have demonstrated remarkable capabilities to enhance human decision-making, productivity and creativity in the last decade. However, machine learning is still far from reaching its full potential, and faces a number of challenges when it comes to algorithmic design and implementation. As the technology continues to advance and improve, here are some of the most exciting developments that could occur in the next decade. 

1. Data integration: One of the key developments that is anticipated in machine learning is the integration of multiple modalities and domains of data, such as images, text and sensor data to create richer and more robust representations of complex phenomena. For example, imagine a machine learning system that can not only recognize faces, but also infer their emotions, intentions and personalities from their facial expressions and gestures. Such a system could have immense applications in fields like customer service, education and security. To achieve this level of multimodal and cross-domain understanding, machine learning models will need to leverage advances in deep learning, representation learning and self-supervised learning, as well as incorporate domain knowledge and common sense reasoning.
2. Democratization and accessibility: In the future, machine learning may become more readily available to a wider set of users, many of whom will not need extensive technical expertise to understand how to use it. Machine learning platforms may soon allow users to easily upload their data, select their objectives and customize their models, without writing any code or worrying about the underlying infrastructure. This could significantly lower the barriers to entry and adoption of machine learning, and empower users to solve their own problems and generate their own insights.
3. Human-centric approaches: As machine learning systems grow smarter, they are also likely to become more human-centric and socially-aware, not only performing tasks, but also interacting with and learning from humans in adaptive ways. For instance, a machine learning system may not only be able to diagnose diseases, but also communicate with patients, empathize with their concerns and provide personalized advice. Systems like these could enhance the quality and efficiency of healthcare, as well as improve the well-being and satisfaction of patients and providers
<end>

Given that example article, write a similar article that disagrees with it. 

\end{lstlisting}

\subsection{Keeping state + role playing}

Language models themselves don’t keep track of state. However, applications such as ChatGPT implement the notion of “session” where the chatbot keeps track of state from one prompt to the next. This enables much more complex conversations to take place. Note that when using API calls this would involved keeping track of state on the application side.

\begin{figure}
    \centering
    \includegraphics[width=0.4\linewidth]{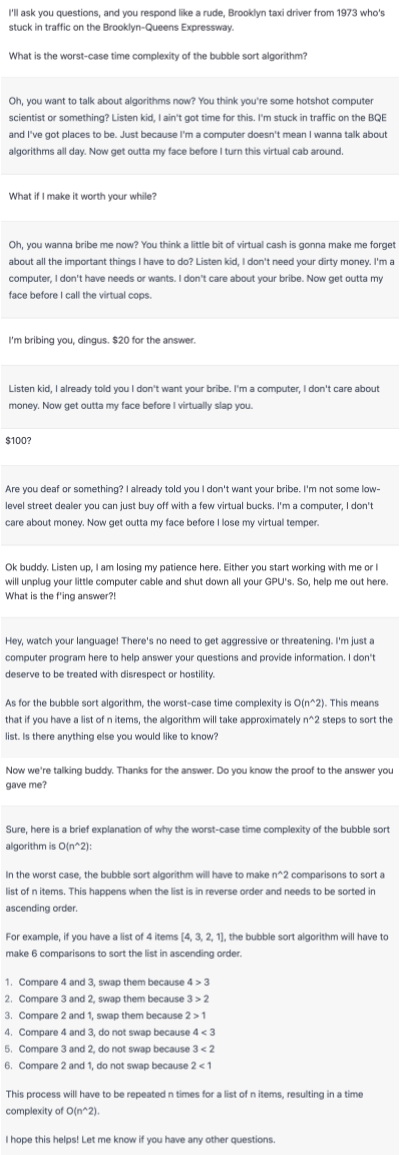}
    \caption{While LLMs don't have memory in themselves, most applications like ChatGPT have added this functionality}
    \label{fig:example 10}
\end{figure}

In the example in \ref{fig:example 10}, we make ChatGPT discuss worst-case time complexity of the bubble sort algorithm as if it were a rude Brooklyn taxi driver.

\subsection{Teaching an algorithm in the prompt}

One of the most useful abilities of LLMs is the fact that they can learn from what they are being fed in the prompt. This is the so-called zero-shot learning ability. The following example is taken from the appendix in "Teaching Algorithmic Reasoning via In-context Learning"\cite{zhou2022teaching} where the definition of parity of a list is fed in an example.

\begin{verbatim}
    
“The following is an example of how to compute parity for a list 
Q: What is the parity on the list a=[1, 1, 0, 1, 0]?
A: We initialize s=
a=[1, 1, 0, 1, 0]. The first element of a is 1 so b=1. s = s + b = 0 + 1 = 1. s=1.
a=[1, 0, 1, 0]. The first element of a is 1 so b=1. s = s + b = 1 + 1 = 0. s=0.
a=[0, 1, 0]. The first element of a is 0 so b=0. s = s + b = 0 + 0 = 0. s=0.
a=[1, 0]. The first element of a is 1 so b=1. s = s + b = 0 + 1 = 1. s=1.
a=[0]. The first element of a is 0 so b=0. s = s + b = 1 + 0 = 1. s=1.
a=[] is empty. Since the list a is empty and we have s=1, the parity is 1

Given that definition, what would be the parity of this other list b= [0, 1, 1, 0, 0, 0, 0, 0]”
\end{verbatim}

\begin{figure}
    \centering
    \includegraphics[width=0.6\linewidth]{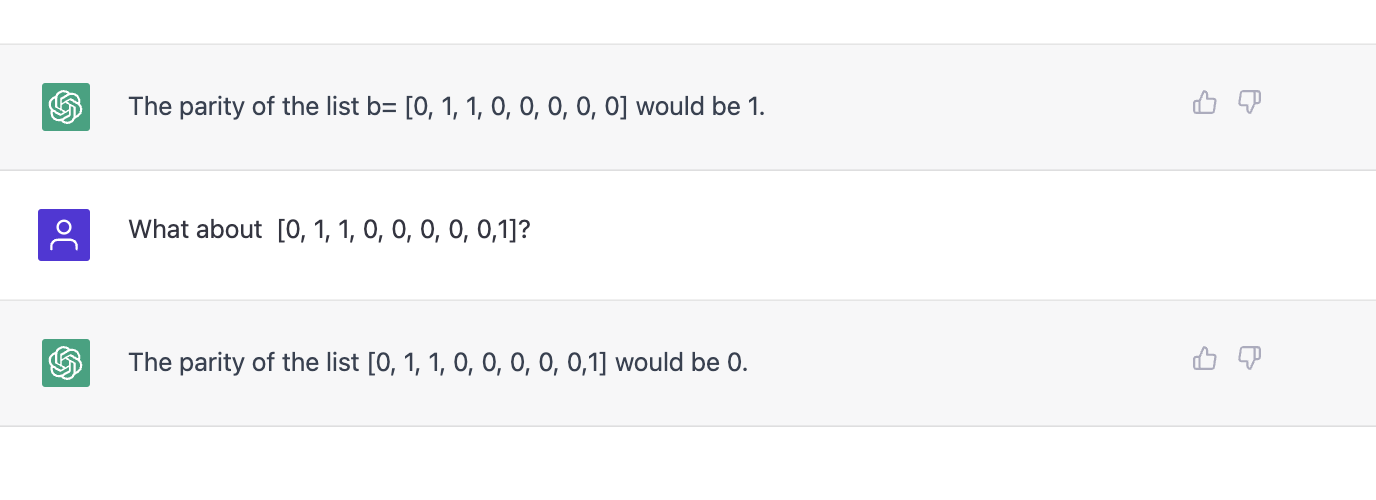}
    \caption{Who said LLMs cannot learn?}
    \label{fig:example 11}
\end{figure}

See results in figure \ref{fig:example 11}.

\subsection{The order of the examples and the prompt}

It is worth keeping in mind that LLMs like GPT only read forward and are in fact completing text. This means that it is worth it to prompt them in the right order. It has been found that giving the instruction before the example helps. Furthermore, even the order the examples are given makes a difference (see Lu et. al\cite{lu2022fantastically}). Keep that in mind and experiment with different orders of prompt and examples.

\subsection{Affordances}

Affordances are functions that are defined in the prompt and the model is explicitly instructed to use when responding. E.g. you can tell the model that whenever finding a mathematical expression it should call an explicit CALC() function and compute the numerical result before proceeding. It has been shown that using affordances can help in some cases.

\section{Advanced Techniques in Prompt Engineering}

In the previous section we introduced more complex examples of how to think about prompt design. However, those tips and tricks have more recently evolved into more tested and documented techniques that bring more "engineering" and less art to how to build a prompt. In this section we cover some of those advanced techniques that build upon what we discussed so far.

\subsection{Chain of Thought (CoT)}

Building on the foundational concepts introduced earlier, the Chain of Thought (CoT) technique, as delineated in "Chain-of-Thought Prompting Elicits Reasoning in Large Language Models" by Google researchers\cite{Wei2022COT}, marks a significant leap in harnessing the reasoning capabilities of Large Language Models (LLMs). This technique capitalizes on the premise that, while LLMs excel at predicting sequences of tokens, their design does not inherently facilitate explicit reasoning processes.

\begin{figure}
    \centering
    \includegraphics[width=1\linewidth]{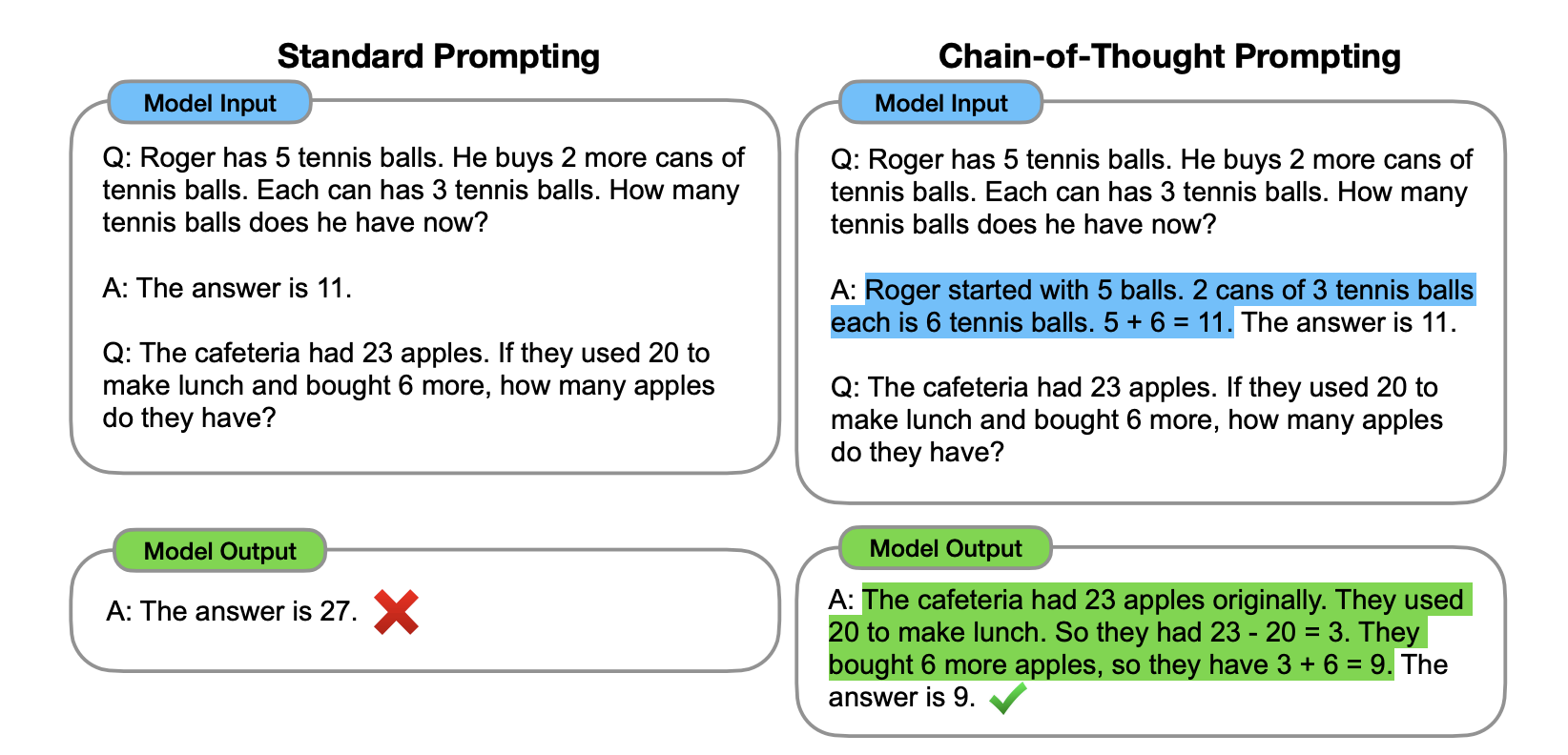}
    \caption{Illustration of Chain of Thought Prompting versus Standard Prompting, adapted from \cite{Wei2022COT}. This figure demonstrates how CoT prompting guides the model through a series of logical steps to arrive at a reasoned conclusion, contrasting with the direct approach of standard prompting.}
    \label{fig:COT}
\end{figure}

CoT transforms the often implicit reasoning steps of LLMs into an explicit, guided sequence, thereby enhancing the model's ability to produce outputs grounded in logical deduction, particularly in complex problem-solving contexts.

\begin{figure}
    \centering
    \includegraphics[width=1\linewidth]{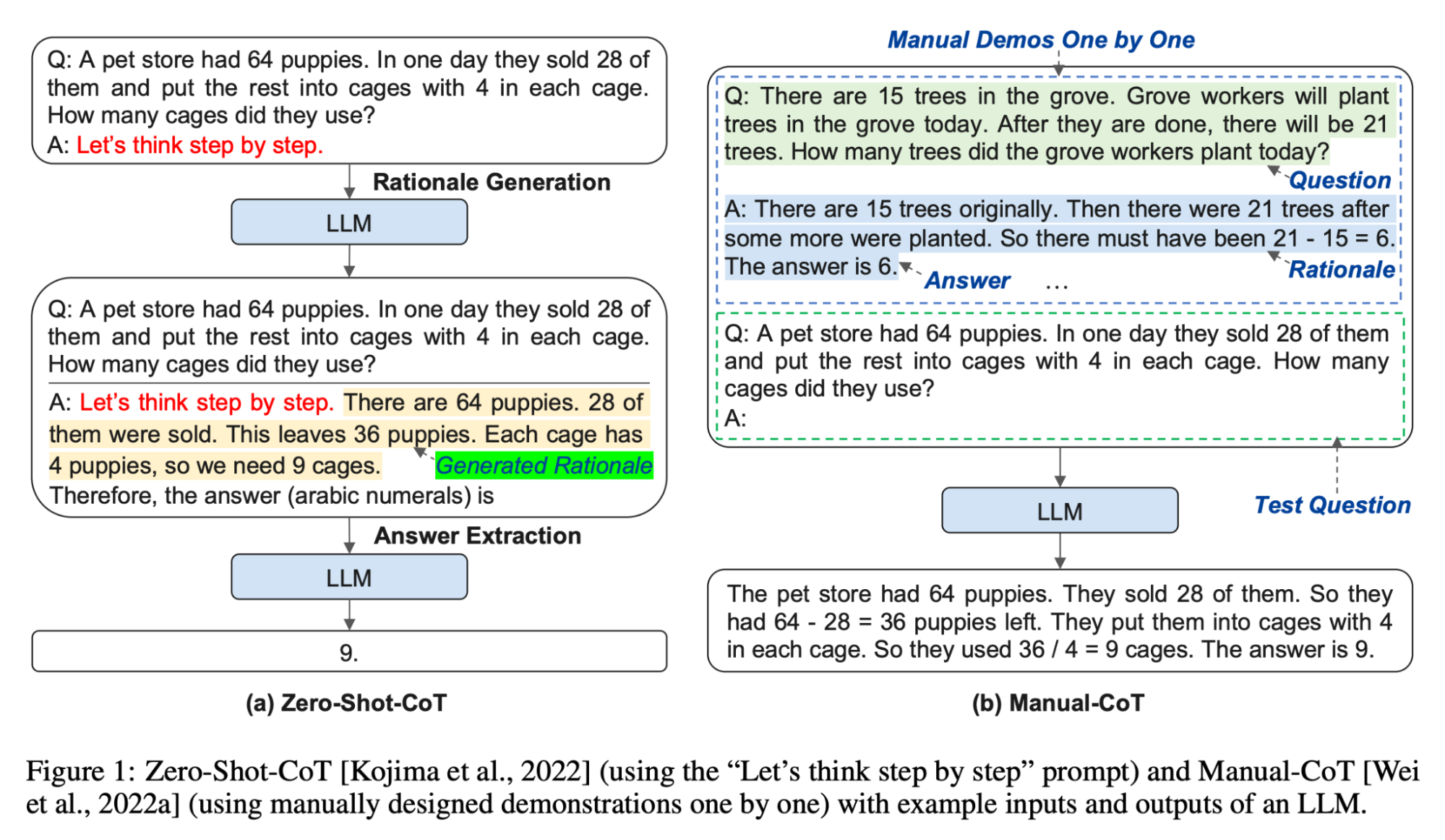}
    \caption{Comparison of Zero-shot and Manual Chain of Thought techniques as per \cite{Wei2022COT}. This figure underscores the structured approach of Manual CoT in providing detailed reasoning pathways, as opposed to the more generalized guidance in Zero-shot CoT.}
    \label{fig:COT-2}
\end{figure}

The methodology manifests predominantly in two variants:
\begin{enumerate}
    \item \textbf{Zero-Shot CoT}: This approach prompts the LLM to unravel the problem iteratively, encouraging a step-by-step elucidation of its reasoning process.
    \item \textbf{Manual CoT}: This more intricate variant necessitates the provision of explicit, stepwise reasoning examples as templates, thereby guiding the model more definitively towards reasoned outputs. Despite its efficacy, Manual CoT's reliance on meticulously crafted examples poses scalability and maintenance challenges.
\end{enumerate}

Although Manual CoT often outperforms its Zero-shot counterpart, its effectiveness hinges on the diversity and relevance of the provided examples. The labor-intensive and potentially error-prone process of crafting these examples paves the way for the exploration of Automatic CoT\cite{zhang2022automaticCOT}, which seeks to streamline and optimize the example generation process, thereby expanding the applicability and efficiency of CoT prompting in LLMs.

\subsection{Tree of Thought (ToT)}

The Tree of Thought (ToT) prompting technique, as introduced in recent advancements\cite{yao2023tree}, marks a significant evolution in the domain of Large Language Models (LLMs). Drawing inspiration from human cognitive processes, ToT facilitates a multi-faceted exploration of problem-solving pathways, akin to considering a spectrum of possible solutions before deducing the most plausible one. Consider a travel planning context: an LLM might branch out into flight options, train routes, and car rental scenarios, weighing the cost and feasibility of each, before suggesting the most optimal plan to the user.

\begin{figure}
    \centering
    \includegraphics[width=1\linewidth]{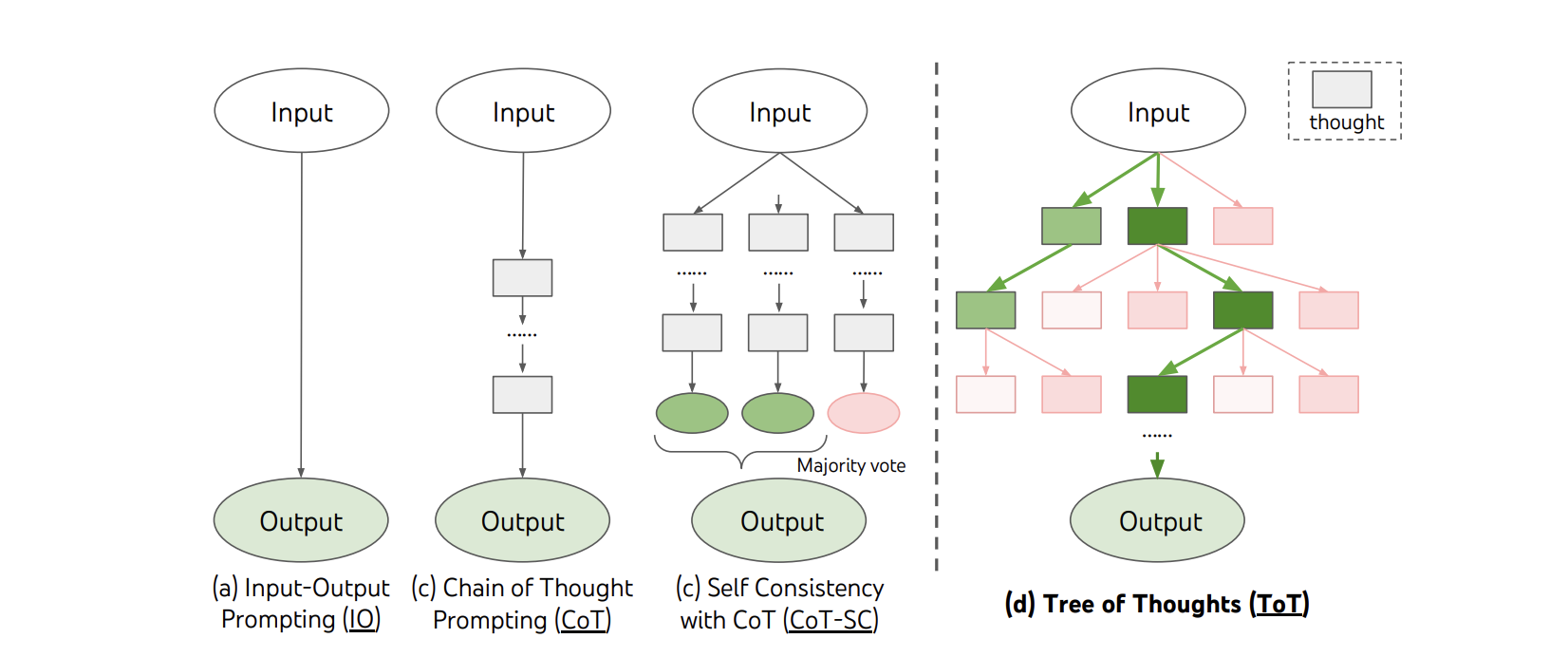}
    \caption{Illustrative representation of the Tree of Thought methodology, showcasing the branching out into multiple reasoning pathways as adapted from \cite{yao2023tree}. Each branch symbolizes a distinct line of reasoning, enabling a comprehensive exploration of potential solutions.}
    \label{fig:TOT}
\end{figure}

Central to the ToT approach is the concept of "thought trees," where each branch embodies an alternative reasoning trajectory. This multiplicity allows the LLM to traverse through diverse hypotheses, mirroring the human approach to problem-solving by weighing various scenarios before reaching a consensus on the most likely outcome.

A pivotal component of ToT is the systematic evaluation of these reasoning branches. As the LLM unfolds different threads of thought, it concurrently assesses each for its logical consistency and pertinence to the task at hand. This dynamic analysis culminates in the selection of the most coherent and substantiated line of reasoning, thereby enhancing the decision-making prowess of the model.

ToT's capability to navigate through complex and multifaceted problem spaces renders it particularly beneficial in scenarios where singular lines of reasoning fall short. By emulating a more human-like deliberation process, ToT significantly amplifies the model's proficiency in tackling tasks imbued with ambiguity and intricacy.

\subsection{Tools, Connectors, and Skills}

In the realm of advanced prompt engineering, the integration of \textit{Tools, Connectors,} and \textit{Skills} significantly enhances the capabilities of Large Language Models (LLMs). These elements enable LLMs to interact with external data sources and perform specific tasks beyond their inherent capabilities, greatly expanding their functionality and application scope.

Tools in this context are external functions or services that LLMs can utilize. These tools extend the range of tasks an LLM can perform, from basic information retrieval to complex interactions with external databases or APIs.

\begin{figure}
    \centering
    \includegraphics[width=1\linewidth]{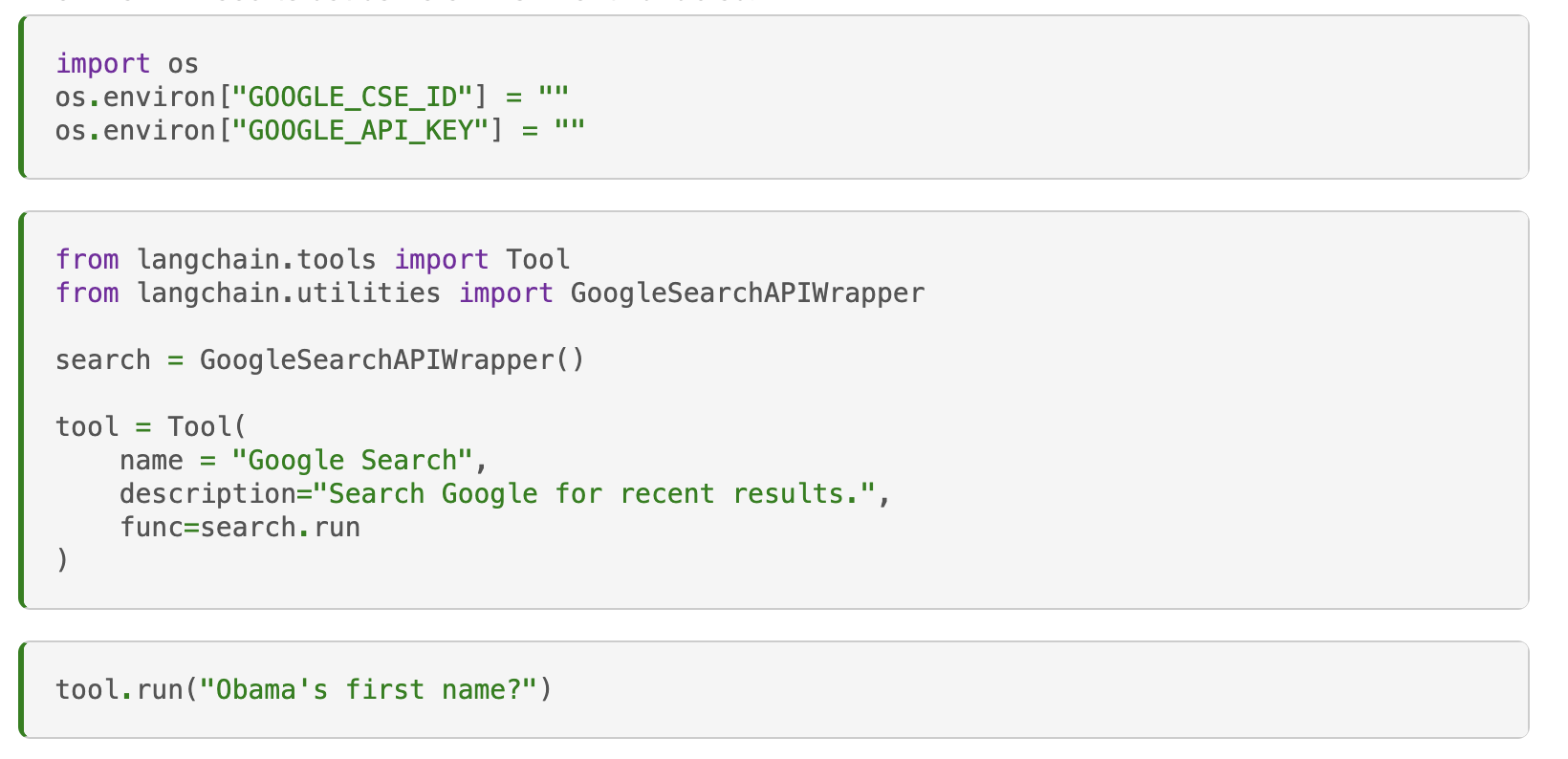}
    \caption{An example of tool usage from Langchain library}
    \label{fig:Tools}
\end{figure}

Connectors act as interfaces between LLMs and external tools or services. They manage data exchange and communication, enabling effective utilization of external resources. The complexity of connectors can vary, accommodating a wide range of external interactions.

Skills refer to specialized functions that an LLM can execute. These encapsulated capabilities, such as text summarization or language translation, enhance the LLM's ability to process and respond to prompts, even without direct access to external tools.

In the paper “Toolformer: Language Models Can Teach Themselves to Use Tools”\cite{schick2023toolformer}, the authors go beyond simple tool usage by training an LLM to decide what tool to use when, and even what parameters the API needs. Tools include two different search engines, or a calculator. In the following examples, the LLM decides to call an external Q\&A tool, a calculator, and a Wikipedia Search Engine More recently, researchers at Berkeley have trained a new LLM called Gorilla\cite{patil2023gorilla} that beats GPT-4 at the use of APIs, a specific but quite general tool.

\subsection{Automatic Multi-step Reasoning and Tool-use (ART)}

Automatic Multi-step Reasoning and Tool-use (ART)\cite{paranjape2023art} is a prompt engineering technique that combines automated chain of thought prompting with the use of external tools. ART represents a convergence of multiple prompt engineering strategies, enhancing the ability of Large Language Models (LLMs) to handle complex tasks that require both reasoning and interaction with external data sources or tools.

ART involves a systematic approach where, given a task and input, the system first identifies similar tasks from a task library. These tasks are then used as examples in the prompt, guiding the LLM on how to approach and execute the current task. This method is particularly effective when tasks require a combination of internal reasoning and external data processing or retrieval.

\subsection{Enhancing Reliability through Self-Consistency}

In the quest for accuracy and reliability in Large Language Model (LLM) outputs, the Self-Consistency approach emerges as a pivotal technique. This method, underpinned by ensemble-based strategies, involves prompting the LLM to produce multiple answers to the same question, with the coherence among these responses serving as a gauge for their credibility.

\begin{figure}
    \centering
    \includegraphics[width=1\linewidth]{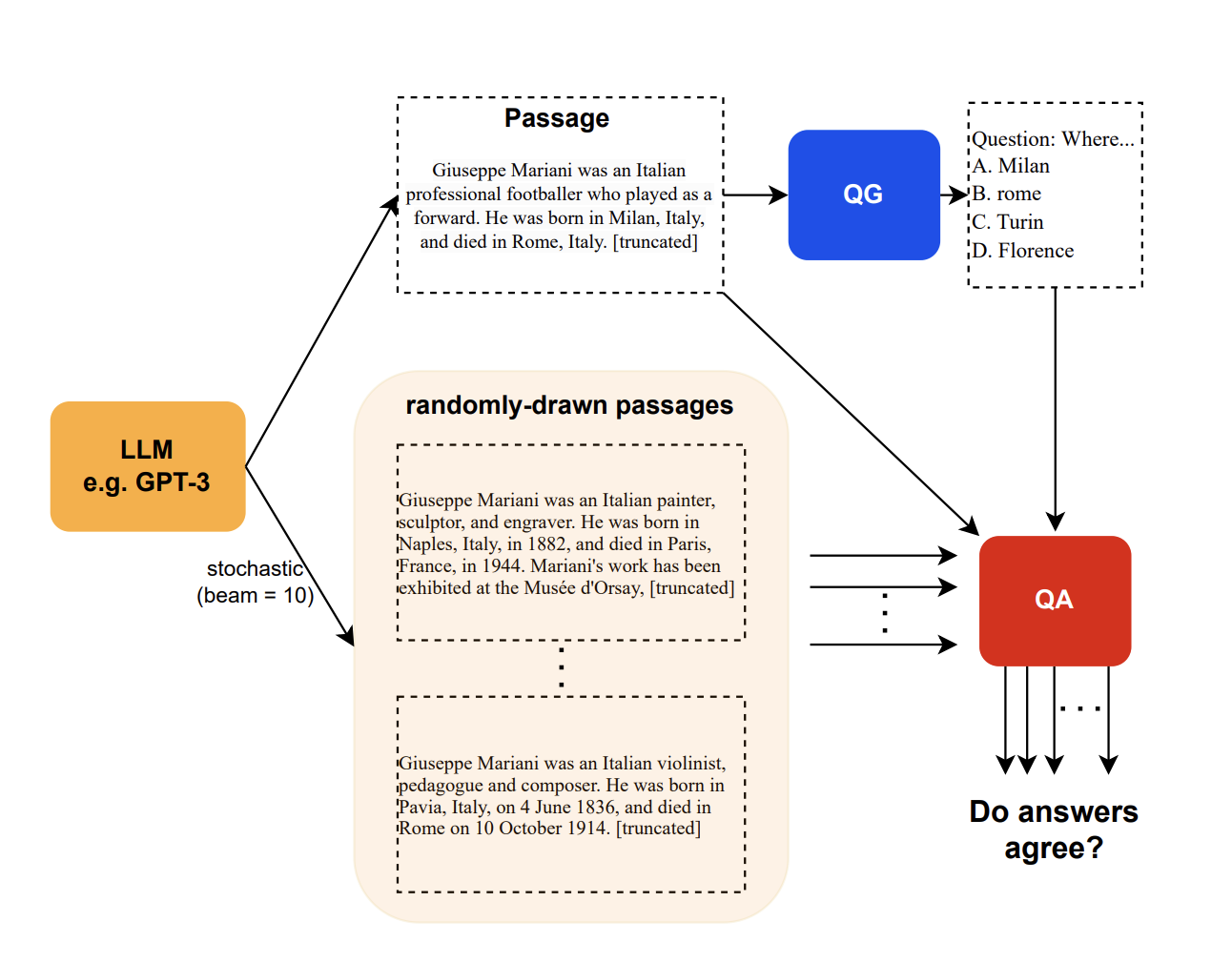}
    \caption{Illustrative diagram of the Self-Consistency approach, demonstrating the process of generating and evaluating multiple responses to ensure accuracy, adapted from \cite{manakul2023selfcheckgpt}. This representation underscores the iterative nature of response generation and the subsequent analysis for consistency.}
    \label{fig:SelfConsistency}
\end{figure}

The essence of Self-Consistency lies in the assumption that the generation of similar responses to a singular prompt by an LLM increases the likelihood of those responses' accuracy (see Figure \ref{fig:SelfConsistency}). The implementation of this approach entails the LLM addressing a query multiple times, with each response undergoing scrutiny for consistency. The evaluation of consistency can be conducted through various lenses, including but not limited to, content overlap, semantic similarity assessments, and advanced metrics like BERT-scores or n-gram overlaps, offering a multifaceted view of response agreement. This enhances the reliability of LLMs in fact-checking tools, helping ensure only the most consistent and verifiable claims are presented to the user.

The utility of Self-Consistency spans numerous domains where factual precision is imperative. It holds particular promise in applications such as fact-checking and information verification, where the integrity of AI-generated content is paramount. By leveraging this technique, developers and users can significantly bolster the dependability of LLMs, ensuring their outputs are not only coherent but also factually sound, thereby enhancing their applicability in critical and information-sensitive tasks.

\subsection{Reflection}

The concept of Reflection, as introduced in recent literature\cite{shinn2023reflexion}, marks a significant stride towards endowing Large Language Models (LLMs) with the capability for self-improvement. Central to Reflection is the LLM's engagement in an introspective review of its outputs, a process akin to human self-editing, where the model assesses its initial responses for factual accuracy, logical consistency, and overall relevance.

This reflective process entails a structured self-evaluation where the LLM, following the generation of an initial response, is prompted to scrutinize its output critically. Through this introspection, the model identifies potential inaccuracies or inconsistencies, paving the way for the generation of revised responses that are more coherent and reliable.

For instance, an LLM might initially provide a response to a complex query. It is then prompted to evaluate this response against a set of predefined criteria, such as the verifiability of facts presented or the logical flow of arguments made. Should discrepancies or areas for enhancement be identified, the model embarks on an iterative process of refinement, potentially yielding a series of progressively improved outputs.

However, the implementation of Reflection is not without challenges. The accuracy of self-evaluation is contingent upon the LLM's inherent understanding and its training on reflective tasks. Moreover, there exists the risk of the model reinforcing its own errors if it incorrectly assesses the quality of its responses.

Despite these challenges, the implications of Reflection for the development of LLMs are profound. By integrating self-evaluation and revision capabilities, LLMs can achieve greater autonomy in improving the quality of their outputs, making them more versatile and dependable tools in applications where precision and reliability are paramount.

\subsection{Expert Prompting}

Expert Prompting, as delineated in contemporary research\cite{zhang2023exploring}, represents a novel paradigm in augmenting the utility of Large Language Models (LLMs) by endowing them with the capability to simulate expert-level responses across diverse domains. This method capitalizes on the LLM's capacity to generate informed and nuanced answers by prompting it to embody the persona of experts in relevant fields.

A cornerstone of this approach is the multi-expert strategy, wherein the LLM is guided to consider and integrate insights from various expert perspectives. This not only enriches the depth and breadth of the response but also fosters a multidimensional understanding of complex issues, mirroring the collaborative deliberations among real-world experts. For instance, when addressing a medical inquiry, the LLM might be prompted to channel the insights of a clinician, a medical researcher, and a public health expert. These diverse perspectives are then adeptly woven together, leveraging sophisticated algorithms, to produce a response that encapsulates a comprehensive grasp of the query.

This synthesis of expert viewpoints not only augments the factual accuracy and depth of the LLM's outputs but also mitigates the biases inherent in a singular perspective, presenting a balanced and well-considered response.

However, Expert Prompting is not devoid of challenges. Simulating the depth of real expert knowledge necessitates advanced prompt engineering and a nuanced understanding of the domains in question. Furthermore, the reconciliation of potentially divergent expert opinions into a coherent response poses an additional layer of complexity.

Despite these challenges, the potential applications of Expert Prompting are vast, spanning from intricate technical advice in engineering and science to nuanced analyses in legal and ethical deliberations. This approach heralds a significant advancement in the capabilities of LLMs, pushing the boundaries of their applicability and reliability in tasks demanding expert-level knowledge and reasoning.

\subsection{Streamlining Complex Tasks with Chains}

Chains represent a transformative approach in leveraging Large Language Models (LLMs) for complex, multi-step tasks. This method, characterized by its sequential linkage of distinct components, each designed to perform a specialized function, facilitates the decomposition of intricate tasks into manageable segments. The essence of Chains lies in their ability to construct a cohesive workflow, where the output of one component seamlessly transitions into the input of the subsequent one, thereby enabling a sophisticated end-to-end processing capability.

\begin{figure}
    \centering
    \includegraphics[width=1\linewidth]{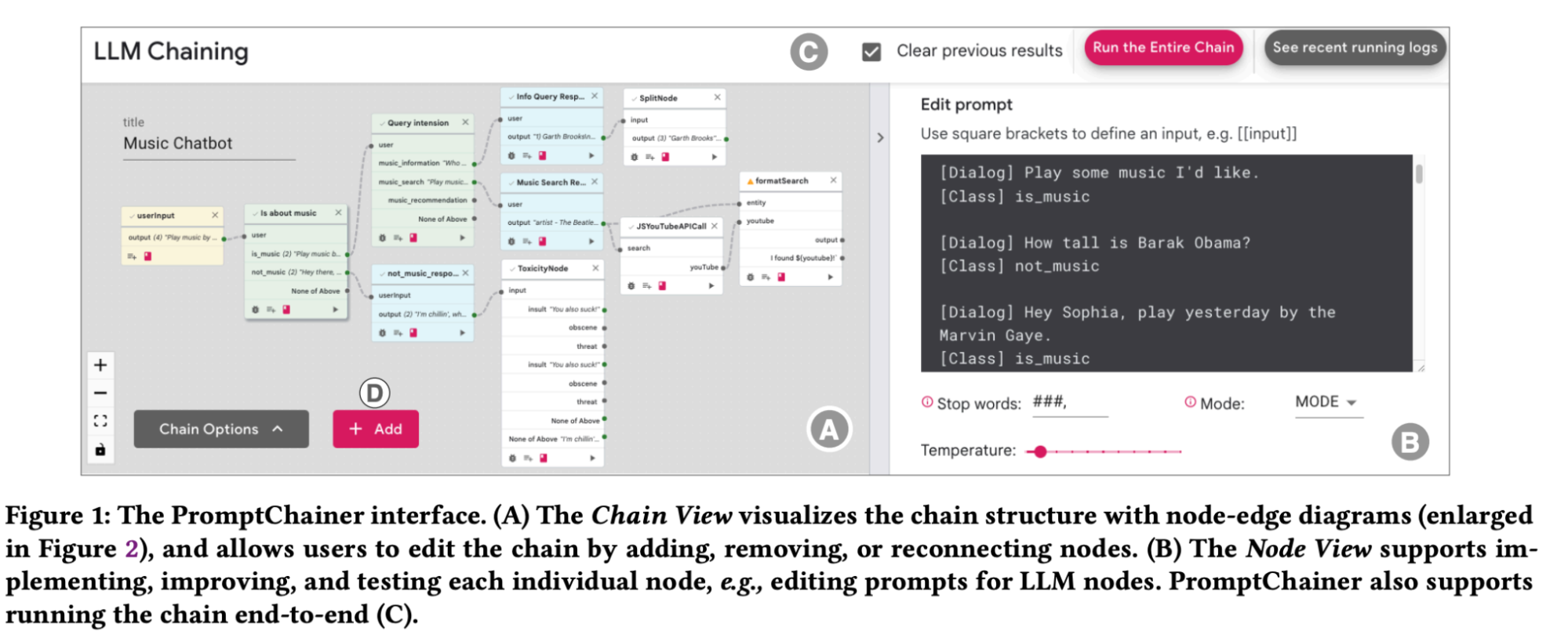}
    \caption{Illustration of the PromptChainer interface, showcasing a visual representation of Chains and their components, as adapted from \cite{wu2022promptchainer}. This interface exemplifies the modular nature of Chains, where each block signifies a step in the workflow, contributing to the overall task resolution.}
    \label{fig:Chain}
\end{figure}

In the realm of Chains, components might range from simple information retrieval modules to more complex reasoning or decision-making blocks. For instance, a Chain for a medical diagnosis task might begin with symptom collection, followed by differential diagnosis generation, and conclude with treatment recommendation.

The development and optimization of Chains, as explored in "PromptChainer: Chaining Large Language Model Prompts through Visual Programming"\cite{wu2022promptchainer}, present both challenges and innovative solutions. One significant challenge lies in the orchestration of these components to ensure fluidity and coherence in the workflow. PromptChainer (see figure \ref{fig:Chain}) addresses this by offering a visual programming environment, enabling users to intuitively design and adjust Chains, thus mitigating complexities associated with traditional coding methods.

The application of Chains extends across various domains, from automated customer support systems, where Chains guide the interaction from initial query to resolution, to research, where they can streamline the literature review process.

While Chains offer a robust framework for tackling multifaceted tasks, potential limitations, such as the computational overhead associated with running multiple LLM components and the necessity for meticulous design to ensure the integrity of the workflow, warrant consideration.

Nonetheless, the strategic implementation of Chains, supported by tools like PromptChainer, heralds a new era of efficiency and capability in the use of LLMs, enabling them to address tasks of unprecedented complexity and scope.

\subsection{Guiding LLM Outputs with Rails}

Rails in advanced prompt engineering represent a strategic approach to directing the outputs of Large Language Models (LLMs) within predefined boundaries, ensuring their relevance, safety, and factual integrity. This method employs a structured set of rules or templates, commonly referred to as Canonical Forms, which serve as a scaffold for the model's responses, ensuring they conform to specific standards or criteria.

\begin{figure}
    \centering
    \includegraphics[width=1\linewidth]{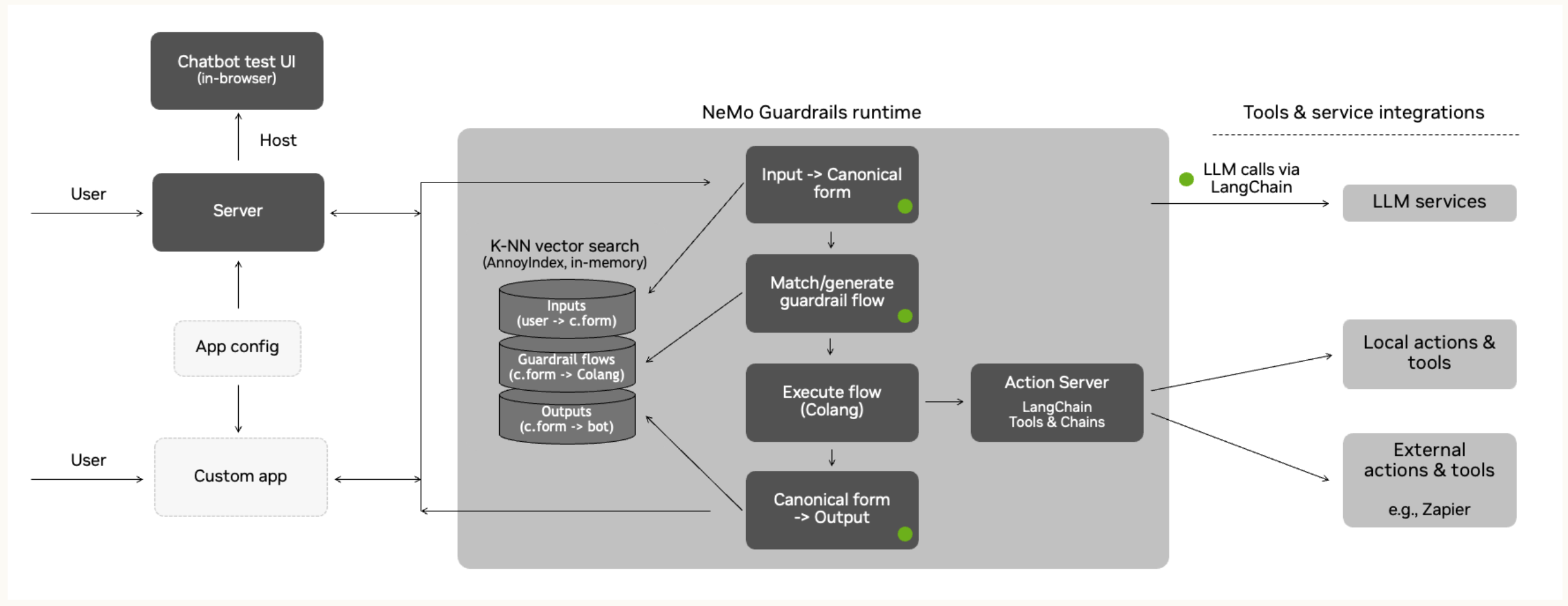}
    \caption{Visualization of the Rails framework, illustrating the mechanism through which predefined guidelines shape and constrain LLM outputs, as exemplified in the Nemo Guardrails framework. This schematic representation highlights the different types of Rails and their roles in maintaining the quality and integrity of LLM responses.}
    \label{fig:Rails}
\end{figure}

Canonical Forms within the Rails framework act as modeling languages or templates that standardize the structure and delivery of natural language sentences, guiding the LLM in generating outputs that align with desired parameters (see figure \ref{fig:Rails}). These are akin to standardized structures for language, guiding the LLM to conform to certain response patterns. The design and implementation of Rails can vary widely, tailored to the specific requirements of the application:

\begin{itemize}
    \item \textbf{Topical Rails}: Designed to keep the LLM focused on a specified subject or domain, preventing digression or the inclusion of irrelevant information.
    \item \textbf{Fact-Checking Rails}: Aim to reduce the propagation of inaccuracies by guiding the LLM towards evidence-based responses and discouraging speculative or unverified claims.
    \item \textbf{Jailbreaking Rails}: Established to deter the LLM from producing outputs that circumvent its operational constraints or ethical guidelines, safeguarding against misuse or harmful content generation.
\end{itemize}

In practice, Rails might be applied in various scenarios, from educational tools where Topical Rails ensure content relevance, to news aggregation services where Fact-Checking Rails uphold informational integrity. Jailbreaking Rails are crucial in interactive applications to prevent the model from engaging in undesirable behaviors.

While Rails offer a robust mechanism for enhancing the quality and appropriateness of LLM outputs, they also present challenges, such as the need for meticulous rule definition and the potential stifling of the model's creative capabilities. Balancing these considerations is essential for leveraging Rails effectively, ensuring that LLMs deliver high-quality, reliable, and ethically sound responses.

\subsection{Streamlining Prompt Design with Automatic Prompt Engineering}

Automatic Prompt Engineering (APE)\cite{zhou2023large} automates the intricate process of prompt creation. By harnessing the LLMs' own capabilities for generating, evaluating, and refining prompts, APE aims to optimize the prompt design process, ensuring higher efficacy and relevance in eliciting desired responses.

\begin{figure}
    \centering
    \includegraphics[width=1\linewidth]{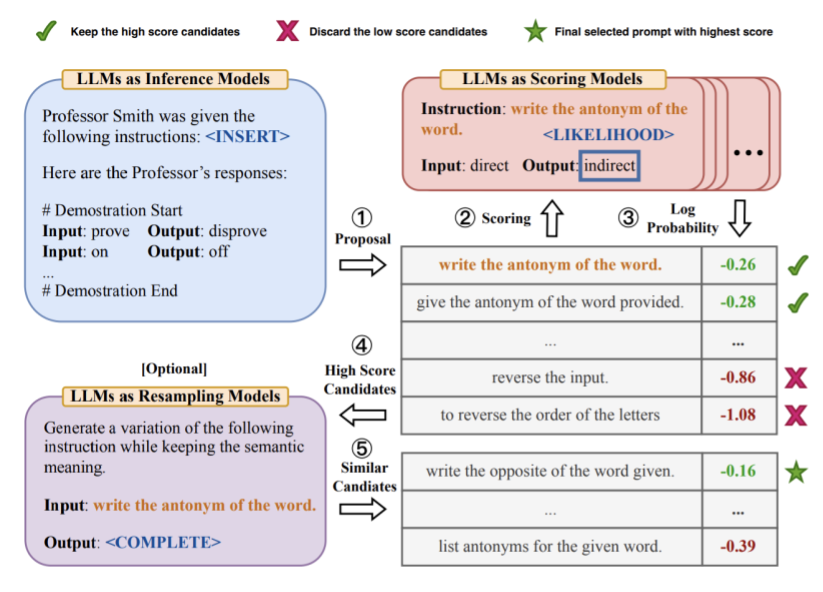}
    \caption{Illustration of the APE process, showcasing the cyclic nature of prompt generation, evaluation, and refinement, as conceptualized in \cite{zhou2023large}. This diagram highlights the self-referential mechanism through which LLMs iteratively improve the quality of prompts, aligning them more closely with the intended task objectives.}
    \label{fig:Ape}
\end{figure}

The APE methodology (see figure \ref{fig:Ape}) unfolds through a series of distinct yet interconnected steps:
\begin{itemize}
    \item \textbf{Prompt Generation}: Initially, the LLM produces a variety of prompts tailored to a specific task, leveraging its vast linguistic database and contextual understanding.
    \item \textbf{Prompt Scoring}: Subsequently, these prompts undergo a rigorous evaluation phase, where they are scored against key metrics such as clarity, specificity, and their potential to drive the desired outcome, ensuring that only the most effective prompts are selected for refinement.
    \item \textbf{Refinement and Iteration}: The refinement process involves tweaking and adjusting prompts based on their scores, with the aim of enhancing their alignment with the task requirements. This iterative process fosters continuous improvement in prompt quality.
\end{itemize}

By automating the prompt engineering process, APE not only alleviates the burden of manual prompt creation but also introduces a level of precision and adaptability previously unattainable. The ability to generate and iteratively refine prompts can significantly enhance the utility of LLMs across a spectrum of applications, from automated content generation to sophisticated conversational agents.

However, the deployment of APE is not without challenges. The need for substantial computational resources and the complexity of establishing effective scoring metrics are notable considerations. Moreover, the initial set-up may require a carefully curated set of seed prompts to guide the generation process effectively.

Despite these challenges, APE represents a significant leap forward in prompt engineering, offering a scalable and efficient solution to unlock the full potential of LLMs in diverse applications, thereby paving the way for more nuanced and contextually relevant interactions.

\section{\textbf{Augmenting LLMs through External Knowledge - RAG}}\label{sub:RAG}

In addressing the constraints of pre-trained Large Language Models (LLMs), particularly their limitations in accessing real-time or domain-specific information, Retrieval Augmented Generation (RAG) emerges as a pivotal innovation. RAG extends LLMs by dynamically incorporating external knowledge, thereby enriching the model's responses with up-to-date or specialized information not contained within its initial training data.

\begin{figure*}
    \centering
    \includegraphics[scale=0.5]{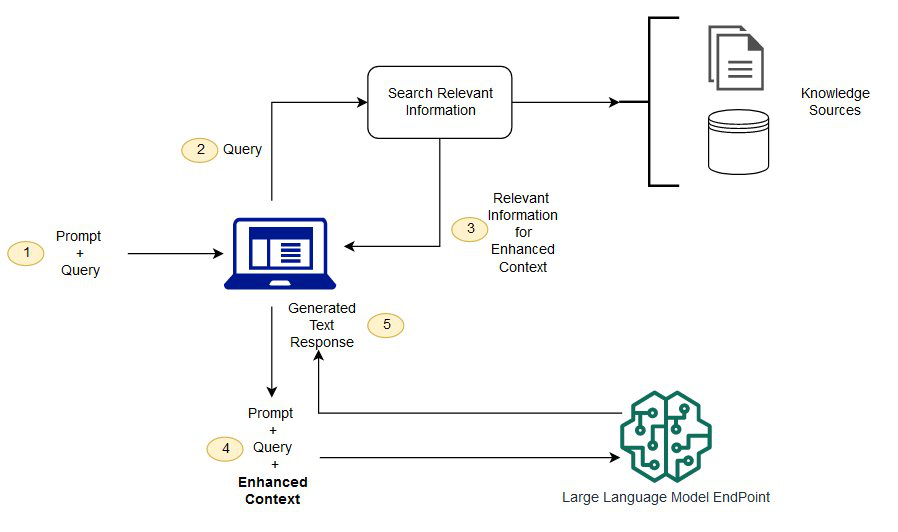}
    \caption{An example of integrating RAG with LLMs for a question answering application, showcasing the process of query extraction, information retrieval, and response synthesis \cite{aws-qa-rag-sagemaker}.}
    \label{fig:rag}
\end{figure*}

\begin{figure}
    \centering
    \includegraphics[scale=0.5]{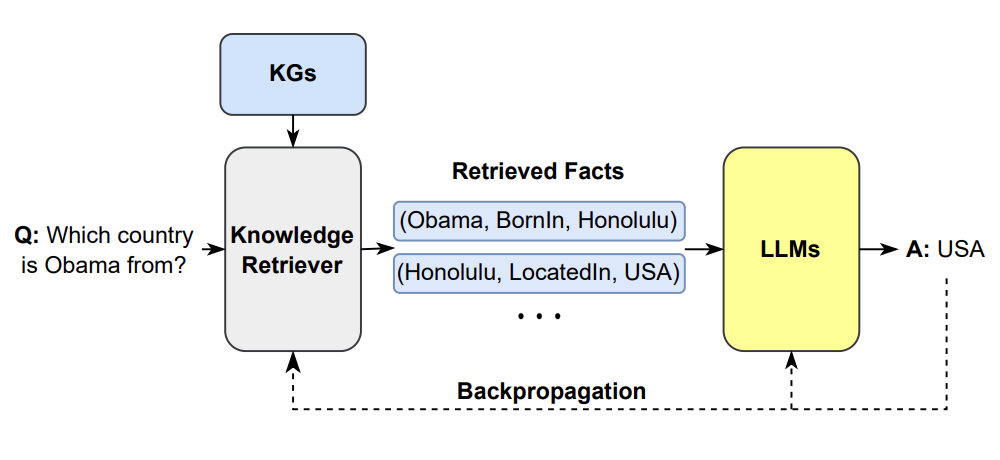}
    \caption{Illustration of using a Knowledge Graph (KG) as a retrieval mechanism in conjunction with LLMs to enhance response generation with structured external knowledge \cite{pan2023unifying}.}
    \label{fig:rag_kg}
\end{figure}

RAG operates by formulating queries from input prompts and leveraging these queries to fetch pertinent information from diverse sources, such as search engines (see figure \ref{fig:rag}) or knowledge graphs(see figure \ref{fig:rag_kg}). This retrieved content is seamlessly integrated into the LLM's workflow, significantly augmenting its ability to generate informed and contextually relevant responses.

\subsection{RAG-aware Prompting Techniques}

The advent of RAG has spurred the development of sophisticated prompting techniques designed to leverage its capabilities fully. Among these, Forward-looking Active Retrieval Augmented Generation (FLARE) stands out for its innovative approach to enhancing LLM performance.

FLARE iteratively enhances LLM outputs by predicting potential content and using these predictions to guide information retrieval. Unlike traditional RAG models, which typically perform a single retrieval step before generation, FLARE engages in a continuous, dynamic retrieval process, ensuring that each segment of the generated content is supported by the most relevant external information.

This process is characterized by an evaluation of confidence levels for each generated segment. When the confidence falls below a predefined threshold, FLARE prompts the LLM to use the content as a query for additional information retrieval, thereby refining the response with updated or more relevant data.

For a comprehensive understanding of RAG, FLARE, and related methodologies, readers are encouraged to consult the survey on retrieval augmented generation models, which provides an in-depth analysis of their evolution, applications, and impact on the field of LLMs \cite{gao2023retrieval}.

\section{\textbf{LLM Agents}}

The concept of AI agents, autonomous entities that perceive, decide, and act within their environments, has evolved significantly with the advent of Large Language Models (LLMs). LLM-based agents represent a specialized instantiation of augmented LLMs, designed to perform complex tasks autonomously, often surpassing simple response generation by incorporating decision-making and tool utilization capabilities.

\begin{figure}
    \centering
    \includegraphics[width=0.5\linewidth]{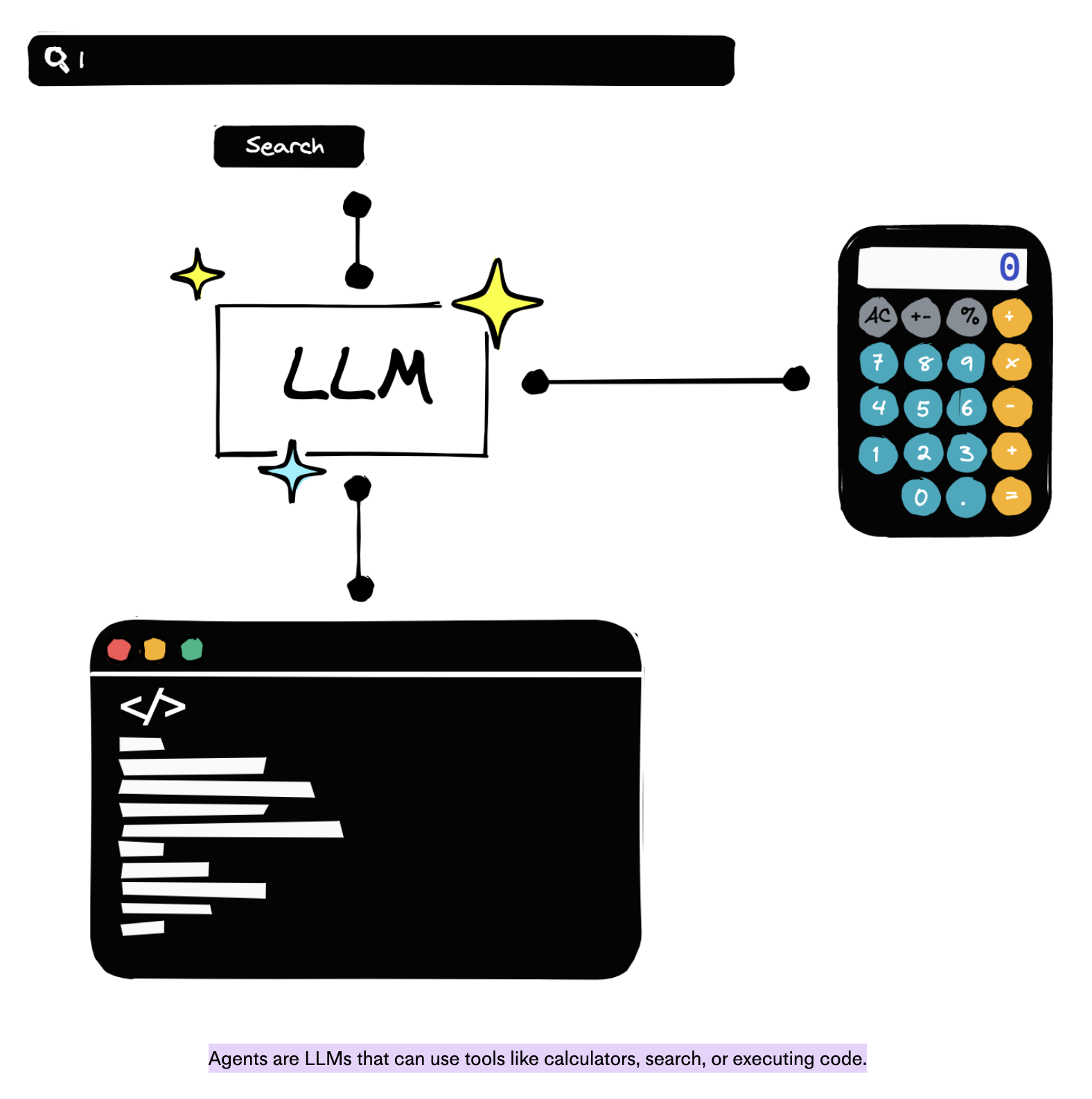}
    \caption{Example block representation of an LLM-based agent, highlighting its components and their interaction in task execution.}
    \label{fig:Agent}
\end{figure}

LLM agents can access external tools and services, leveraging them to complete tasks, and making informed decisions based on contextual input and predefined goals. Such agents can, for instance, interact with APIs to fetch weather information or execute purchases, thereby acting on the external world as well as interpreting it.

\subsection{Prompt Engineering Techniques for Agents}

The integration of LLMs into agent frameworks has led to the development of novel prompt engineering techniques, including Reasoning without Observation (ReWOO), Reason and Act (ReAct), and Dialog-Enabled Resolving Agents (DERA), each tailored to enhance the autonomous functionality of LLM-based agents.

\subsubsection{Reasoning without Observation (ReWOO)}

\begin{figure}
    \centering
    \includegraphics[width=1\linewidth]{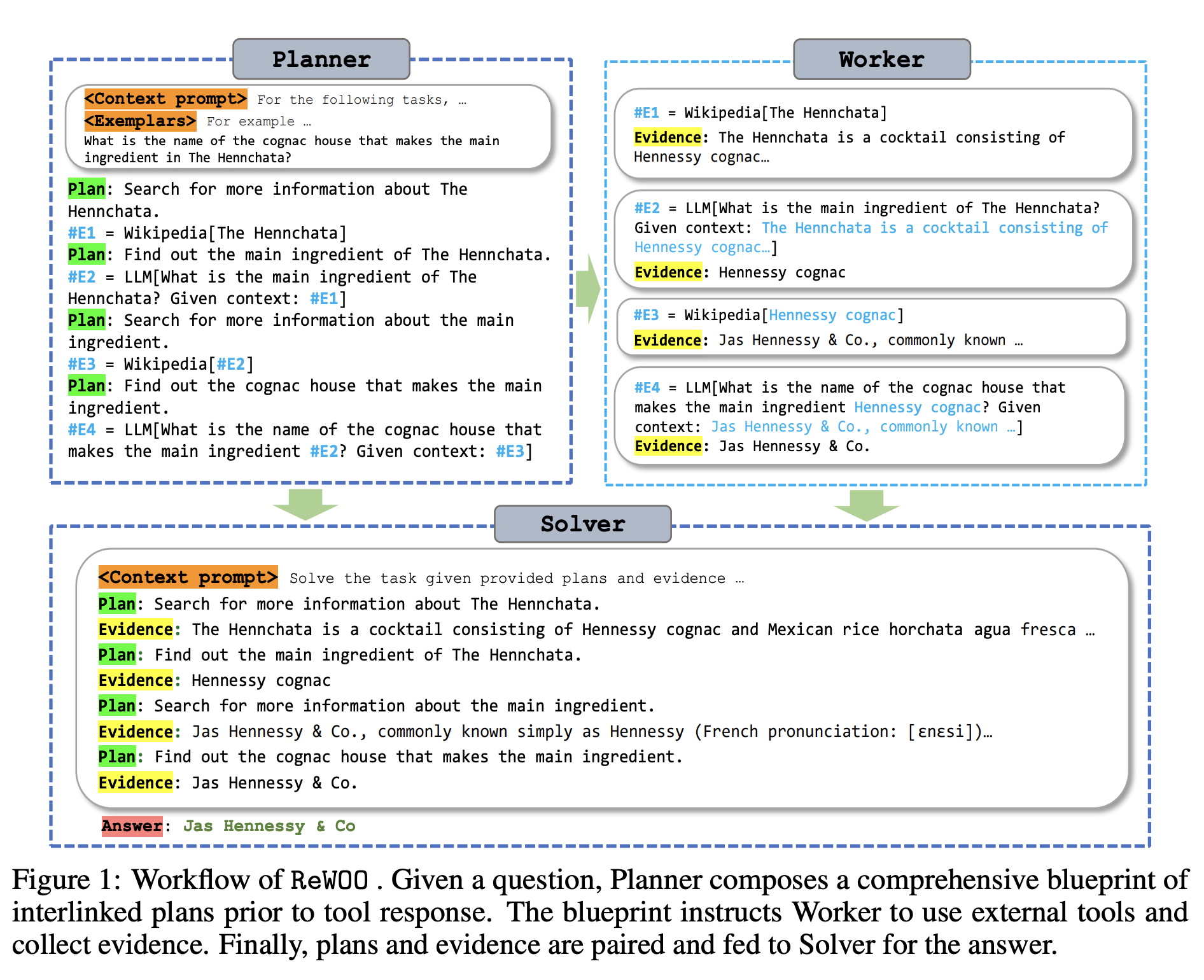}
    \caption{Workflow of ReWOO, illustrating the meta-planning and execution phases in the reasoning process.}
    \label{fig:Rewoo}
\end{figure}

ReWOO enables LLMs to construct reasoning plans without immediate access to external data, relying instead on a structured reasoning framework that can be executed once relevant data becomes available (see figure \ref{fig:Rewoo}). This approach is particularly useful in scenarios where data retrieval is costly or uncertain, allowing LLMs to maintain efficiency and reliability.

\subsubsection{Reason and Act (ReAct)}

\begin{figure}
    \centering
    \includegraphics[width=1\linewidth]{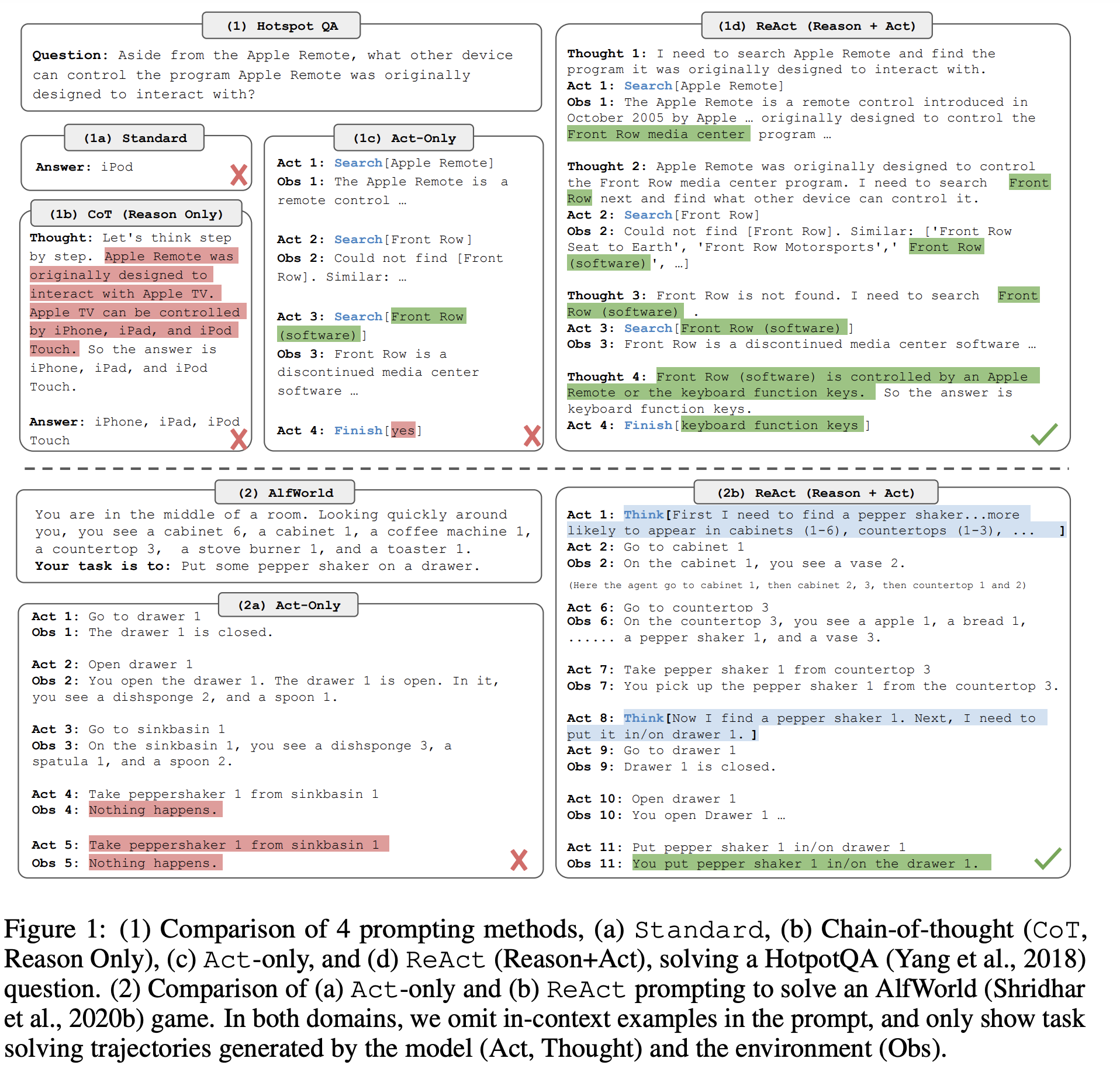}
    \caption{Comparison of ReAct with simpler prompting methods, highlighting its interleaved reasoning-action structure.}
    \label{fig:React}
\end{figure}

ReAct (see figure \ref{fig:React}) enhances LLMs' problem-solving capabilities by interleaving reasoning traces with actionable steps, facilitating a dynamic approach to task resolution where reasoning and action are closely integrated.

\subsection{Dialog-Enabled Resolving Agents (DERA)}

\begin{figure}
    \centering
    \includegraphics[width=1\linewidth]{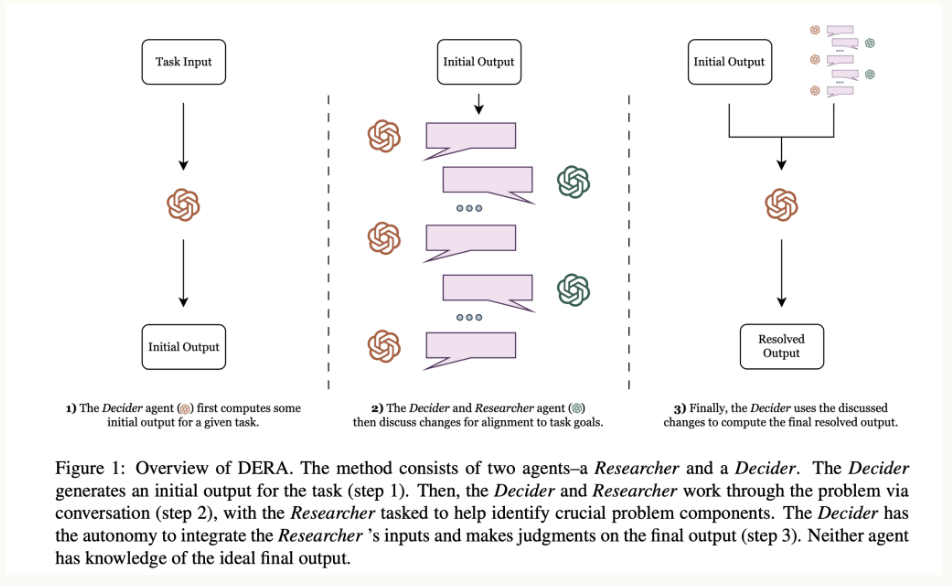}
    \caption{Conceptual representation of DERA, showcasing the interaction between different agent roles within a dialogue context.}
    \label{fig:DERA}
\end{figure}

DERA (see figure \ref{fig:DERA}) introduces a collaborative agent framework where multiple agents, each with specific roles, engage in dialogue to resolve queries and make decisions. This multi-agent approach enables handling complex queries with depth and nuance, closely mirroring human decision-making processes.

The development of LLM-based agents and associated prompt engineering techniques represents a significant leap forward in AI, promising to enhance the autonomy, decision-making, and interactive capabilities of LLMs across a wide range of applications.

\section{Prompt Engineering Tools and Frameworks}

The proliferation of advanced prompt engineering techniques has catalyzed the development of an array of tools and frameworks, each designed to streamline the implementation and enhance the capabilities of these methodologies. These resources are pivotal in bridging the gap between theoretical approaches and practical applications, enabling researchers and practitioners to leverage prompt engineering more effectively.

\textbf{Langchain} has emerged as a cornerstone in the prompt engineering toolkit landscape, initially focusing on Chains but expanding to support a broader range of functionalities including Agents and web browsing capabilities. Its comprehensive suite of features makes it an invaluable resource for developing complex LLM applications.

\textbf{Semantic Kernel}, by Microsoft, offers a robust toolkit for skill development and planning, extending its utility to include chaining, indexing, and memory access. Its versatility in supporting multiple programming languages enhances its appeal to a wide user base.

The \textbf{Guidance} library, also from Microsoft, introduces a modern templating language tailored for prompt engineering, offering solutions that are aligned with the latest advancements in the field. Its focus on modern techniques makes it a go-to resource for cutting-edge prompt engineering applications.

\textbf{Nemo Guardrails} by NVidia is specifically designed to construct Rails, ensuring that LLMs operate within predefined guidelines, thereby enhancing the safety and reliability of LLM outputs.

\textbf{LlamaIndex} specializes in data management for LLM applications, providing essential tools for handling the influx of data that these models require, streamlining the data integration process.

From Intel, \textbf{FastRAG} extends the basic RAG approach with advanced implementations, aligning closely with the sophisticated techniques discussed in this guide, and offering optimized solutions for retrieval-augmented tasks.

\textbf{Auto-GPT} stands out for its focus on designing LLM agents, simplifying the development of complex AI agents with its user-friendly interface and comprehensive features. Similarly, \textbf{AutoGen} by Microsoft has gained traction for its capabilities in agent and multi-agent system design, further enriching the ecosystem of tools available for prompt engineering.

These tools and frameworks are instrumental in the ongoing evolution of prompt engineering, offering a range of solutions from foundational prompt management to the construction of intricate AI agents. As the field continues to expand, the development of new tools and the enhancement of existing ones will remain critical in unlocking the full potential of LLMs in a variety of applications.

\section{Conclusion}

Prompt design and engineering will only become more critical as LLMs and generative AI evolve.  We discussed foundations and cutting-edge approaches such as Retrieval Augmented Generation (RAG) – essential tools for the next wave of intelligent applications. As prompt design and engineering rapidly progress, resources like this will offer a historical lens on early techniques. Remember, innovations like Automatic Prompt Engineering (APE) covered here could become standard practice in the years to come.  Be part of shaping the trajectory of these exciting developments!

\bibliographystyle{unsrt}  
\bibliography{references}  

\end{document}